\documentclass[aps, prd, onecolumn, tightenlines, notitlepage, superscriptaddress, nofootinbib, preprintnumbers, floatfix,showkeys,11pt,altaffilletter]{revtex4-2}
\usepackage{upgreek}

\usepackage[official]{eurosym}
\usepackage[normalem]{ulem}
\usepackage{amstext}

\usepackage{graphicx}
\graphicspath{{figures/}}
\usepackage{url}
\usepackage{color}
\usepackage{ulem}
\usepackage[version=4]{mhchem}
\usepackage[utf8]{inputenc}
\usepackage{fontawesome}
\usepackage{yfonts}
\usepackage{mathrsfs}

\pdfoutput=1
\usepackage{textcomp}
\usepackage{comment}
\usepackage{yfonts}
\usepackage{epsfig,amsfonts,mathrsfs,graphicx,color,slashed,multirow}
\usepackage{latexsym,graphicx,slashed,color,enumerate,url,cancel,gensymb}
\usepackage{textcomp}

\usepackage[x11names]{xcolor}
\usepackage[colorlinks,pdfstartview=FitV,breaklinks=true]{hyperref}
\usepackage{booktabs}
\usepackage{adjustbox}
\usepackage{textgreek} 
\usepackage{caption, subcaption} 
\usepackage{booktabs} 
\usepackage{float}

\usepackage{lmodern}
\usepackage{ae,aecompl}
\usepackage{appendix}
\usepackage{orcidlink}
\usepackage{nccmath} 
\usepackage{textcomp}
\allowdisplaybreaks 


\def\cevns{CE\textnu NS}

\def\d{\mathrm{d}}
\newcommand{\qtransfer}{\left|\mathbf{q}\right|}

\definecolor{vdrgreen}{rgb}{0.0, 0.6, 0.0}

\definecolor{magenta(dye)}{rgb}{0.79, 0.08, 0.48}
\definecolor{byzantium}{rgb}{0.44, 0.16, 0.39}

\def\rd#1{\textcolor{red}{#1}}

\DeclareMathOperator{\diag}{diag}

\makeatletter
    \newcommand{\colorboxed}[3][white]{\fcolorbox{#2}{#1}{\m@th$\displaystyle#3$}}
\makeatother

\AtBeginDocument{\hypersetup{citecolor=magenta(dye),linkcolor=magenta(dye),urlcolor=magenta(dye)}}
\usepackage{appendix}

\begin{document}

\title{{\LARGE Testing light and heavy vector mediators\\ with solar \cevns~measurements}}

\author{Valentina De Romeri~\orcidlink{0000-0003-3585-7437}}
\email{deromeri@ific.uv.es}
\affiliation{Instituto de F\'{i}sica Corpuscular (IFIC), CSIC‐Universitat de Val\'encia, E-46980 Valencia, Spain}

\author{Dimitrios K. Papoulias~\orcidlink{0000-0003-0453-8492}}\email{dimitrios.papoulias@uni-hamburg.de}
\affiliation{Institute of Experimental Physics, University of Hamburg, 22761, Hamburg, Germany}

\author{Federica Pompa~\orcidlink{0000-0002-9591-8361}}
\email{federica.pompa@lngs.infn.it}
\affiliation{Istituto Nazionale di Fisica Nucleare (INFN), Laboratori Nazionali del Gran Sasso, 67100 Assergi, L’Aquila (AQ), Italy
}

\author{Gonzalo Sanchez Garcia~\orcidlink{0000-0003-1830-2325}}%
\email{g.sanchez@ciencias.unam.mx}%
\affiliation{Departamento de F\'{i}sica, Facultad de Ciencias, Universidad Nacional Aut\'onoma de M\'exico,
Apartado Postal 70-542, Ciudad de M\'exico 04510, M\'exico}%

\author{Christoph A. Ternes~\orcidlink{0000-0002-7190-1581}}
\email{christoph.ternes@lngs.infn.it}
\affiliation{Gran Sasso Science Institute, Viale F. Crispi 7, L’Aquila, 67100, Italy}
\affiliation{Istituto Nazionale di Fisica Nucleare (INFN), Laboratori Nazionali del Gran Sasso, 67100 Assergi, L’Aquila (AQ), Italy
}


\begin{abstract}

The recent observation of coherent elastic neutrino–nucleus scattering from solar $^8$B neutrinos in dark matter direct detection experiments has inaugurated the \emph{neutrino fog} era, highlighting the extended potential of these experiments as precision neutrino observatories. Recent measurements by the XENONnT, PandaX-4T, and LUX-ZEPLIN experiments provide new opportunities to test Standard Model predictions and to probe  physics beyond it, in complementarity with dedicated neutrino facilities. 
We perform a combined analysis of nuclear recoil data from these three facilities to extract information on the solar $^8$B neutrino flux normalization and on the weak mixing angle at low-momentum transfer. We further investigate the impact of new vector interactions on the solar neutrino event rate, deriving constraints on nonstandard neutrino interactions and on scenarios with light vector mediators. Our results demonstrate that dark matter detectors are rapidly becoming complementary to terrestrial neutrino experiments in probing neutrino interactions, and already set competitive bounds on both light and heavy vector mediators.
 
\end{abstract}
\maketitle

\section{Introduction}
Coherent elastic neutrino–nucleus scattering (\cevns) has recently emerged as a powerful tool for precision tests of the Standard Model (SM) and for searches of physics beyond it. In this neutral-current process, a low-energy neutrino transfers momentum coherently to an entire nucleus~\cite{Abdullah:2022zue}. Although theoretically predicted several decades ago~\cite{Freedman:1973yd}, its observation remained experimentally challenging for a long time. The same coherence effect that enhances the SM cross section, scaling approximately with the square of the  number of neutrons in the nuclear target, implies nuclear recoils at the sub-keV level~\cite{Drukier:1984vhf}, requiring detectors with extremely low energy thresholds and ultra-low backgrounds. The first measurement of this process was achieved by the COHERENT collaboration~\cite{COHERENT:2017ipa} using pion-decay-at-rest neutrinos at a spallation source. Subsequent observations with different targets~\cite{COHERENT:2020iec,COHERENT:2021xmm,Adamski:2024yqt} and reactor neutrinos~\cite{Colaresi:2022obx,Ackermann:2025obx} have firmly established the process and confirmed its coherence properties.

Remarkably, the experimental conditions required to observe \cevns~closely match those of modern dark matter (DM) direct detection experiments~\cite{Goodman:1984dc,Schumann:2019eaa,Billard:2021uyg}. Large target masses, low recoil thresholds, and strong background suppression define both research programs. Ton-scale dual-phase liquid xenon detectors such as XENONnT~\cite{XENON:2023cxc}, LUX-ZEPLIN (LZ)~\cite{LZ:2022lsv}, and PandaX-4T~\cite{PandaX:2024qfu} have reached unprecedented sensitivities in their primary search for weakly interacting massive particles. Although no conclusive DM signal has been observed, the continuous reduction of energy thresholds has opened sensitivity to astrophysical neutrinos. In particular, high-energy solar $^8$B neutrinos can now induce detectable \cevns~events in these detectors. This development marks the practical onset of the so-called \textit{neutrino fog}~\cite{OHare:2021utq,Monroe:2007xp,Vergados:2008jp,Strigari:2009bq,Billard:2013qya,DeRomeri:2025nkx}, in which neutrino-induced nuclear recoils constitute an irreducible background for DM searches. At the same time, it transforms DM facilities into precision neutrino observatories, probing neutrino properties in a regime complementary to dedicated neutrino experiments (see, e.g.,~\cite{Harnik:2012ni,AtzoriCorona:2022jeb,deGouvea:2021ymm,Giunti:2023yha,Cerdeno:2016sfi,Dutta:2017nht,Gelmini:2018gqa,Essig:2018tss,Boehm:2020ltd,AristizabalSierra:2020edu,AristizabalSierra:2020zod,Amaral:2020tga,Dutta:2020che,Suliga:2020jfa,Amaral:2021rzw,DeRomeri:2024dbv,Aalbers:2022dzr,Alonso-Gonzalez:2023tgm,Amaral:2023tbs,Majumdar:2024dms,DeRomeri:2024hvc,DeRomeri:2025nkx,Gehrlein:2025isp,Blanco-Mas:2024ale,AtzoriCorona:2025gyz,Demirci:2023tui,AtzoriCorona:2025xwr}).

This transition has now materialized experimentally. With their latest exposures, the XENONnT \cite{XENON:2024ijk,XENON:2026ydt}, PandaX-4T~\cite{PandaX:2024muv}, and LZ~\cite{LZ:2025igz} collaborations have reported the first indications of nuclear recoils consistent with \cevns~from solar $^8$B neutrinos. In the case of XENONnT and PandaX-4T, the background-only hypothesis is disfavored at around three standard deviations, while LZ reports a 4.5$\sigma$ significance. These measurements represent the first evidence of \cevns~on a xenon target and the first observation of solar neutrino–induced nuclear recoils in DM experiments. Interpreted within the SM, the extracted $^8$B flux is consistent with solar model predictions~\cite{Vinyoles:2016djt} and with results from dedicated neutrino observatories~\cite{SNO:2011hxd}. Beyond confirming SM expectations, these results establish DM direct detection experiments as a new precision frontier for neutrino physics. They have already motivated phenomenological studies addressing nonstandard neutrino interactions (NSI)~\cite{AristizabalSierra:2024nwf,Li:2024iij,Gehrlein:2025isp,Li:2026mco}, new light mediators~\cite{Xia:2024ytb,DeRomeri:2024iaw,Blanco-Mas:2024ale,AtzoriCorona:2025gyz}, electromagnetic properties~\cite{DeRomeri:2024hvc}, and the determination of electroweak parameters at low-momentum transfer~\cite{Maity:2024aji,DeRomeri:2024iaw,LZ:2025igz,AtzoriCorona:2025gyz}. As exposures continue to increase, solar \cevns~measurements at DM facilities are expected to play a central role in testing neutrino interactions and physics beyond the SM, as anticipated in earlier studies (see, e.g., Refs.~\cite{AristizabalSierra:2017joc,Cerdeno:2016sfi,Gonzalez-Garcia:2018dep,Amaral:2023tbs,Park:2023hsp,AristizabalSierra:2019ykk,Majumdar:2021vdw,Schwemberger:2022fjl,Bertuzzo:2017tuf,Dutta:2017nht}).

In this work, we investigate the implications of the recent observation of $^8$B solar \cevns\ events in DM direct detection experiments through a combined spectral analysis of the nuclear-recoil data reported by XENONnT~\cite{XENON:2024ijk,XENON:2026ydt}, PandaX-4T~\cite{PandaX:2024muv}, and LZ~\cite{LZ:2025igz}. We first extract determinations of the $^8$B solar neutrino flux normalization and of the weak mixing angle at low-momentum transfer, exploiting the complementarity of the three experiments. We then explore the impact of new neutrino interactions, considering both neutrino NSI described by effective point-like vector operators and scenarios with a light vector mediator. While these possibilities have previously been studied in the context of DM detectors (see, e.g.,~\cite{Dutta:2017nht,AristizabalSierra:2017joc,AristizabalSierra:2019ykk,Dutta:2019oaj,Amaral:2023tbs}), here we update and extend those analyses by confronting them with the latest solar \cevns\ data.
For NSI, we allow for both flavor-conserving and flavor-violating couplings and consistently include matter effects in the Sun, accounting for their impact on neutrino propagation and flavor conversion. For light mediators, instead, we focus on flavor-diagonal scenarios, namely a universal coupling framework and an extension of the SM with a gauged $B-L$ symmetry, so that solar neutrino oscillations remain unaltered and new physics effects arise solely at detection.
We compare the resulting bounds with those obtained from terrestrial \cevns~experiments~\cite{Giunti:2019xpr,Papoulias:2019xaw,Miranda:2020tif,DeRomeri:2022twg,Coloma:2023ixt,Coloma:2022avw,DeRomeri:2025csu} and neutrino oscillation experiments~\cite{Coloma:2023ixt}, demonstrating the complementarity and competitive sensitivity of our present $^8$B solar-neutrino \cevns–based constraints.

The paper is organized as follows. In Sec.~\ref{sec:theory} we review neutrino propagation in the Sun and the \cevns~cross section in the SM as well as how it is modified in the presence of new interactions with heavy or light vector mediators. The statistical treatment of the LZ data is described in Sec.~\ref{sec:analyses}; for the analysis of XENONnT and PandaX-4T data we rely on our previous work~\cite{DeRomeri:2024iaw}. Results are presented in Sec.~\ref{sec:results}, and conclusions are finally given in Sec.~\ref{sec:concl}.

\section{Theoretical framework and cross sections}
\label{sec:theory}

The presence of new vector interactions can modify the detection rate of solar neutrinos in two distinct ways. First, the presence of NSI introduces additional contributions to the matter potential, thus altering neutrino flavor evolution inside the Sun. This modifies the flavor composition of the solar neutrino flux ($\nu_e$, $\nu_\mu$, and $\nu_\tau$) that reaches the detector.
%
Second, the new, possibly flavor-dependent, interactions directly affect detection by contributing to the \cevns~cross section.

In this section, we first discuss how heavy, purely vector mediators modify neutrino oscillation probabilities through their impact on the solar matter potential. We then examine their effect on the weak nuclear charge entering the \cevns~cross section, considering both heavy and light mediators. We emphasize that the light mediator scenarios considered in this work are flavor diagonal and therefore do not alter neutrino flavor conversion in the Sun. 

\subsection{Neutrino nonstandard interactions and light vector mediators}
\label{subsec:NSI}

Neutrino NSI provide a convenient and model-independent framework to parametrize the effects of heavy new physics in neutrino processes at low energies. In particular, heavy vector mediators appearing in many extensions of the SM, such as extra neutral gauge bosons, induce effective dimension-six four-fermion operators that modify neutrino neutral-current (NC) interactions. In this work we focus on operators leading to purely vector interactions.

In the presence of  NSI, the NC Lagrangian is modified as~\cite{Ohlsson:2012kf,Miranda:2015dra,Farzan:2017xzy}
\begin{equation}
\mathcal{L}^{\mathrm{NSI}}_{\mathrm{NC}}
= -2\sqrt{2}\, G_F 
\sum_{q,\ell,\ell',X}
\varepsilon^{qX}_{\ell\ell'}
\left(\bar{\nu}_\ell \gamma^\mu P_L \nu_{\ell'}\right)
\left(\bar{q} \gamma_\mu P_X q\right) \, ,
\label{NSI:lagrangian}
\end{equation}
where $q=\{u,d\}$ denotes first-generation quarks, $\ell,\ell'$ label neutrino flavors, and $P_X$ with $X=L,R$ are the chirality projectors. The dimensionless parameters $\varepsilon^{qX}_{\ell\ell'}$ quantify the strength of NSI relative to the Fermi constant, $G_F$, and they can be either flavor conserving ($\ell=\ell'$, possibly non-universal) or flavor changing ($\ell\neq\ell'$). In this work, we focus on the vector part of the interaction given by $\varepsilon^{qV}_{\ell\ell'} = \varepsilon^{qL}_{\ell\ell'} + \varepsilon^{qR}_{\ell\ell'}$. Note that, in principle, axial-vector combinations can also be formed, according to $\varepsilon^{qA}_{\ell\ell'} = \varepsilon^{qL}_{\ell\ell'} - \varepsilon^{qR}_{\ell\ell'}$. However, since axial-vector interactions do not modify the matter potential in the Sun, and since the axial contribution to the \cevns~cross section is small, we do not consider them in this work. The propagation of neutrinos in a medium is actually only sensitive to the combinations

\begin{equation}
    \varepsilon_{\ell\ell'} = \varepsilon^{eV}_{\ell\ell'} + \frac{N_u}{N_e}\varepsilon^{uV}_{\ell\ell'} + \frac{N_d}{N_e}\varepsilon^{dV}_{\ell\ell'}\,,
\end{equation}
where $\varepsilon^{eV}_{\ell\ell'}$ indicates a nonstandard interaction with electrons while $N_e$, $N_u$ and $N_d$ are the number densities of electrons, $u$-quarks and $d$-quarks, respectively. 
Note also that in this work we do not address NSI with electrons, which have been recently studied in~\cite{Coloma:2023ixt,BOREXINO:2026owb}.
Since in this work we study, in addition to neutrino propagation, effects of NSI on the {\cevns} cross section, we focus only on NSI with $u$- or $d$-quarks. 

Note that  the parametrization in Eq.~\eqref{NSI:lagrangian} is not gauge-invariant\footnote{One possibility, not explored here, is to expand this formalism to all gauge invariant dimension-six operators using an effective field theory approach (see, for instance,~\cite{Breso-Pla:2023tnz,Li:2026mco,Coloma:2024ict}).}.
In general, gauge invariance would imply that NC NSI are accompanied by analogous operators in the charged lepton sector, which are tightly constrained by experimental data~\cite{Gavela:2008ra,Antusch:2008tz}. However, these bounds can be significantly relaxed in new physics scenarios where NSI arise from the exchange of light neutral mediators with masses well below the electroweak scale (see, e.g.,~\cite{Farzan:2015hkd,Farzan:2015doa,Heeck:2018nzc,Babu:2017olk} and the review~\cite{Proceedings:2019qno}). In such cases, large NSI strengths leading to observable effects in current experiments become viable.
Motivated by this model realizations of such new interactions, it seems convenient to extend our analysis to scenarios in which the new vector interaction is mediated by a light particle (with a mass $m_V$ comparable to the typical momentum transfer in this process, $\mathcal{O}\sim 10$ MeV).
In this case, the Lagrangian is given by

\begin{equation}
\mathcal{L} \;\supset\;
- \frac{1}{4} V_{\mu\nu} V^{\mu\nu}
+ \frac{1}{2} m_V^2 V_\mu V^\mu
+ V_\mu \left(
g_{\nu V}\, \bar{\nu}_\alpha \gamma^\mu P_L \nu_\alpha
+ g_{qV}\, \bar{q} \gamma^\mu q
\right) \, ,
\label{eq:Lag_light_vector}
\end{equation}
where the coupling between neutrinos (quarks) and the new mediator $V$ is given by $g_{\nu V}$ $(g_{q V})$, and $V_{\mu\nu} = \partial_\mu V_\nu - \partial_\nu V_\mu$.

\subsection{Neutrino propagation in the Sun}
\label{subsec:nuprop}
In the presence of NSI, neutrino flavor evolution in matter is governed by the effective Hamiltonian

\begin{equation}
    \mathcal{H}_{\text{NSI}} = \mathcal{H}_0
    + \sqrt{2}G_FN_e
    \begin{pmatrix}
        1 & 0 & 0 \\ 0 & 0 & 0 \\ 0 & 0 & 0 \\ 
    \end{pmatrix}
    + \sqrt{2}G_F\sum_qN_q
    \begin{pmatrix}
        \varepsilon^{qV}_{ee} & \varepsilon^{qV}_{e\mu} & \varepsilon^{qV}_{e\tau} \\ 
        \varepsilon^{qV}_{e\mu} & \varepsilon^{qV}_{\mu\mu} & \varepsilon^{qV}_{\mu\tau}\\ 
        \varepsilon^{qV}_{e\tau} & \varepsilon^{qV}_{\mu\tau} & \varepsilon^{qV}_{\tau\tau}\\ 
    \end{pmatrix}\,,
\end{equation}
where $\mathcal{H}_0 = \frac{1}{2E_\nu}U\diag(0,\Delta m_{21}^2,\Delta m_{31}^2) U^\dagger$ is the vacuum Hamiltonian, with $U$ the PMNS mixing matrix, and $E_\nu$ the neutrino energy. The second term corresponds to the standard matter potential, while the third term encodes NC NSI with first-generation quarks ($q=u,d$), weighted by their number densities $N_q$.
In our analysis, the solar oscillation parameters are fixed to the values recently reported by JUNO~\cite{JUNO:2025gmd}. 
The remaining oscillation parameters are taken from Ref.~\cite{deSalas:2020pgw}, except for the atmospheric mixing angle, for which we adopt $\sin^2\theta_{23}=0.5$, consistent with current global fits~\cite{deSalas:2020pgw,Capozzi:2025wyn,Esteban:2024eli}. This choice implies equal muon and tau flavor components in the solar neutrino flux at Earth.
Note that, in principle, the non-diagonal parameters in the matrix in the last term can be complex.  Here, however, we consider only real NSI. 
We have numerically verified that, including CP-phases or simply allowing for the non-diagonal parameters to take negative values produces the same bounds on the NSI parameters. 

For solar neutrinos, adiabatic flavor conversion probabilities can be computed in the mass-dominance limit, $\Delta m_{31}^2 \gg 2\sqrt{2}G_F N_e E_\nu$, where the matter potential is small compared to the atmospheric mass-squared splitting. In this regime, the three-flavor evolution effectively reduces to a two-flavor problem (see, e.g.,~\cite{AristizabalSierra:2017joc,Coloma:2023ixt}). The electron-neutrino survival probability can then be written as~\cite{Kuo:1986sk}
\begin{equation}
    \mathcal{P}_{ee}(E_\nu,r) = \cos^4\theta_{13}~\mathcal{P}_\text{eff}(E_\nu,r)+\sin^4\theta_{13}\,,
\end{equation}
where $r=R/R_\odot$ is the production point inside the Sun in units of the solar radius, $R_\odot$. The effective two-flavor probability is given by the Parke formula~\cite{Parke:1986jy}
\begin{equation}
  \mathcal{P}_\text{eff}(E_\nu,r) = \frac{1}{2} \left[1+\cos(2\theta_M)\cos(2\theta_{12})\right]\,,
\end{equation}
where $\theta_M$ is the effective mixing angle in matter,
\begin{equation}
    \cos(2\theta_M) = \frac{\Delta m_{21}^2\cos(2\theta_{12})-\mathcal{A}}{\sqrt{(\Delta m_{21}^2\cos(2\theta_{12})-\mathcal{A})^2+(\Delta m_{21}^2{ \sin(2\theta_{12})}+\mathcal{B})^2}}\,.
\end{equation}
The NSI effects on the matter potential are present through the parameters $\mathcal{A}$ and $\mathcal{B}$, defined as
\begin{eqnarray}
    \mathcal{A} &=& 4\sqrt{2}E_\nu G_F N_e(r)\left(\frac{\cos^2\theta_{13}}{2}-\sum_q\frac{N_q(r)}{N_e(r)}\varepsilon_D^q\right)\,,\\
    \mathcal{B} &=& 4\sqrt{2}E_\nu G_F \sum_qN_q(r)\varepsilon_N^q\,.
\end{eqnarray}
 The effective NSI combinations entering solar neutrino propagation are given by (with $c_{ij}\equiv\cos\theta_{ij}$ and $s_{ij}\equiv\sin\theta_{ij}$)~\cite{Coloma:2023ixt,Gonzalez-Garcia:2013usa}

\begin{eqnarray}
    \varepsilon_D^q &=& -\frac{c_{13}^2}{2} (\varepsilon_{ee}^{qV} - \varepsilon_{\mu\mu}^{qV}) + \frac{s_{23}^2-s_{13}^2c_{23}^2}{2}(\varepsilon_{\tau\tau}^{qV} - \varepsilon_{\mu\mu}^{qV})\\
    &+& s_{13}c_{13} (s_{23}\varepsilon_{e\mu}^{qV} + c_{23}\varepsilon_{e\tau}^{qV}) - (1 + s_{13}^2)c_{23}s_{23}\varepsilon_{\mu\tau}^{qV}\nonumber\,,\\
    \varepsilon_N^q &=& s_{13} c_{23} s_{23} (\varepsilon_{\tau\tau}^{qV} - \varepsilon_{\mu\mu}^{qV}) + c_{13} (c_{23} \varepsilon_{e\mu}^{qV} - s_{23} \varepsilon_{e\tau}^{qV}) + s_{13}(s_{23}^2 - c_{23}^2) \varepsilon_{\mu\tau}^{qV} \,.
\end{eqnarray}

Finally, the survival probability must be averaged over the spatial distribution of $^8$B production region inside the Sun\footnote{We have verified that the contribution from $hep$ neutrinos is negligible, while all other solar neutrino components do not yield nuclear recoils within our region of interest.}. The oscillation probabilities at Earth are therefore given by
\begin{eqnarray}
    \mathcal{P}_{ee}(E_\nu) &=& \int_0^1 \rho_{\text{B}}(r)~\mathcal{P}_\text{eff}(E_\nu,r)~\d r\,,\\
    \mathcal{P}_{e\mu}(E_\nu) &=& (1-\mathcal{P}_{ee}(E_\nu))\cos^2\theta_{23}\,,\\
    \mathcal{P}_{e\tau}(E_\nu) &=& (1-\mathcal{P}_{ee}(E_\nu))\sin^2\theta_{23}
\end{eqnarray}
where $\rho_{\text{B}}(r)$ denotes the normalized radial distribution of $^8$B neutrino production, satisfying $\int_0^1 \rho_{\text{B}}(r)\,\mathrm{d}r=1$, and is taken from Ref.~\cite{Bahcall:2005va}.
The differential flux of neutrinos of flavor $\nu_\ell$ at the detector is then
\begin{equation}
    \frac{\d\phi_\ell}{\d E_\nu} = \frac{\d\phi_0}{\d E_\nu}~\mathcal{P}_{e\ell}(E_\nu)\,,
\label{eq:osc-flux}
\end{equation}
where $\mathrm{d}\phi_0/\mathrm{d}E_\nu$ is the unoscillated $^8$B solar neutrino flux, taken from Refs.~\cite{bahcall_web,Bahcall:1996qv}, with total normalization fixed to $5.46\times10^6~\mathrm{cm}^{-2}\mathrm{s}^{-1}$ as given in~\cite{Vinyoles:2016djt,Baxter:2021pqo} for the high-metallicity B16-GS98 model.

The approach outlined above is valid only under the assumption of adiabatic neutrino evolution in the interior of the Sun. This condition is always satisfied within the standard three-neutrino oscillation paradigm, but it may be violated in the presence of large NSI. In our analysis, we follow the procedure suggested in Ref.~\cite{Coloma:2023ixt}: we determine the regions in the $(\varepsilon_D^q,\varepsilon_N^q)$ parameter space for NSI with $u$- and $d$-quarks where adiabaticity may fail, and verify whether any combinations of $\varepsilon_{\ell\ell'}^{qV}$ considered in our study lie within those regions. We find that, in our combined analysis, no parameter points lead to non-adiabatic evolution\footnote{For the analysis of PandaX-4T data alone, some parameter points may lie in non-adiabatic regions. However, since the combined results are dominated by LZ data, we consistently adopt the adiabatic approximation throughout.}. Therefore, the use of the adiabatic approach is fully justified in our analysis.

\subsection{\cevns~cross sections}
\subsubsection{\cevns~cross section in the Standard Model}
\label{subsec:CEvNS-SM}

%
Within the SM, the differential cross section for \cevns~with respect to the nuclear recoil energy, $T_{\mathcal N}$, can be written as~\cite{Freedman:1973yd,Barranco:2005yy}
\begin{equation}
\label{eq:xsec_CEvNS_SM}
\left. \frac{\d\sigma_{\nu_\ell \mathcal{N}}}{\d T_{\mathcal N}}\right|_{\rm SM}
= \frac{G_F^2\, m_{\mathcal N}}{\pi}
\left(Q_{V,\ell}^{\rm SM}\right)^2
F_W^2(\qtransfer^2)
\left(1 - \frac{m_{\mathcal N} T_{\mathcal N}}{2 E_\nu^2}\right),
\end{equation}
where $m_{\mathcal N}$ is the nuclear mass. In the previous expression, we neglected terms that are suppressed by $T_{\mathcal N}/m_{\mathcal N}$ and higher-order contributions of $\mathcal{O}(T_{\mathcal N}^2)$.
The quantity $Q_{V,\ell}^{\rm SM}$ represents the SM weak vector charge of the nucleus, defined as
\begin{equation}
\label{eq:CEvNS_SM_QV}
Q_{V,\ell}^{\rm SM} = g_{V,\ell}^p\, Z + g_{V,\ell}^n\, N ,
\end{equation}
with $Z$ and $N$ being the proton and neutron number,
respectively. At tree level, the corresponding vector couplings are given by $g_{V,\ell}^p = (1/2 - 2 \sin^2\theta_W)/2$ for protons and $g_{V,\ell}^n = -1/2$ for neutrons, while in the SM the flavor dependence becomes only relevant beyond tree level, see e.g. Ref.~\cite{Cadeddu:2020lky}.
Through the proton term, the weak charge depends explicitly on the weak mixing angle, whose value at low energies is obtained by renormalization group evolution from the Z-pole value and corresponds to $\sin^2\theta_W = 0.23857(5)$~\cite{ParticleDataGroup:2024cfk}.

The effects of the nuclear target structure are encoded in the weak nuclear form factor $F_W(\qtransfer^2)$, which accounts for the finite spatial extent of the nucleus. For \cevns~induced by $^8$B solar neutrinos, the typical momentum transfer is sufficiently small that the form-factor suppression is modest. 
In our numerical analysis, we adopt the Klein-Nystrand parametrization~\cite{Klein:1999qj}
\begin{equation}
\label{eq:KNFF}
F_W(\qtransfer^2)
= 3\,\frac{j_1(\qtransfer R_A)}{\qtransfer R_A}
  \left(\frac{1}{1 + \qtransfer^2 a_k^2}\right),
\end{equation}
where $j_1(x) = \sin x/x^2 - \cos x/x$ is the spherical Bessel function of first order, $a_k = 0.7~\mathrm{fm}$, and $R_A = 1.23\,A^{1/3}$ (in fm) denotes the nuclear root-mean-square radius, with $A$ the atomic mass number. The momentum transfer relevant for \cevns~can be estimated as $\qtransfer = \sqrt{2 m_{\mathcal N} T_{\mathcal N}}/197.327~\mathrm{fm}^{-1}$, corresponding to a characteristic scale of $\mathcal{O}(10)$~MeV.

\subsubsection{\cevns~cross section with NSI}

NSI lead to a flavor-dependent modification of the SM weak charge defined in Eq.~(\ref{eq:CEvNS_SM_QV}), such that $(Q_{V,\ell}^{\mathrm{SM}})^2 \to (Q_{V,\ell}^{\mathrm{NSI}})^2$ with~\cite{Barranco:2005yy} 
\begin{eqnarray}
\left(Q_{V,\ell}^{\mathrm{NSI}}\right)^2 &=&
\left[
\left(g_{V,\ell}^p + 2\varepsilon^{uV}_{\ell\ell} + \varepsilon^{dV}_{\ell\ell}\right) Z
+
\left(g_{V,\ell}^n + \varepsilon^{uV}_{\ell\ell} + 2\varepsilon^{dV}_{\ell\ell}\right) N
\right]^2
\nonumber\\
&+&
\sum_{\ell'\neq\ell}
\left|
\left(2\varepsilon^{uV}_{\ell\ell'} + \varepsilon^{dV}_{\ell\ell'}\right) Z
+
\left(\varepsilon^{uV}_{\ell\ell'} + 2\varepsilon^{dV}_{\ell\ell'}\right) N
\right|^2\,.
\label{eq:qvNSI}
\end{eqnarray}
For an incoming $\nu_\ell$, the \cevns~cross section must be appropriately weighted by the relevant neutrino energy flux following Eq.~\eqref{eq:osc-flux}.

\subsubsection{\cevns~cross section in presence of new light vector mediators}
\label{subsec:LM}
%
In the presence of a light vector mediator of mass $m_V$, the \cevns~cross section is modified to~\cite{Candela:2024ljb}
\begin{align} \left.\dfrac{\d \sigma_{\nu_\ell \mathcal{N}}}{\d T_\mathcal{N}}\right|_{V} (E_\nu, T_\mathcal{N}) =& \, \left[1+ \kappa \frac{C_V}{\sqrt{2}G_F Q_{V,\ell}^\mathrm{SM}\left(m_{V}^2+2 m_\mathcal{N} T_\mathcal{N}\right)}\right]^2 \left.\frac{\d\sigma_{\nu_\ell \mathcal{N}}}{\d T_\mathcal{N}}\right|_\mathrm{SM} \, , \label{eq:cross-section-vector-CEvNS} \end{align}
where the squared bracket explicitly accounts for the interference between the SM amplitude and the contribution induced by the new interaction.

Assuming universal quark couplings, the effective coupling to the nucleus is encoded in the coefficient $C_V \equiv 3 A g_V^2 = 3 A g_{\nu V} g_{qV}$,
with $g_{\nu V}$ and $g_{qV}$ denoting the fundamental couplings of the vector mediator to neutrinos and quarks, respectively. The factor $\kappa$ accounts for the specific $U(1)'$ charge assignments of the model under consideration.
We restrict ourselves to flavor-diagonal interactions and consider two benchmark scenarios: (i) a phenomenological model with universal couplings, for which $\kappa = 1$, and (ii) a $B-L$ extension of the SM~\cite{Langacker:2008yv,Okada:2018ktp}, characterized by $\kappa = -1/3$. While the universal-coupling scenario is not anomaly-free and is introduced here purely as an effective description, it allows for destructive interference with the SM contribution. In contrast, in the $B-L$ model only constructive interference is possible. 

\section{Data analysis}
\label{sec:analyses}

The analysis of PandaX-4T data follows the procedure described in Refs.~\cite{DeRomeri:2024iaw,DeRomeri:2024hvc}. While XENONnT data~\cite{XENON:2024ijk} were also analyzed in those references, we update our treatment of XENONnT using the more recent dataset~\cite{XENON:2026ydt}, which we now describe in detail.

We analyze each science run (SR) of XENONnT individually and include all relevant observables provided in the data release: the corrected secondary scintillation signal (cS2), the ratio S2$/\Delta t$ where $\Delta t$ is the drift time between the S1 and S2 signals, encoding the depth of the interaction vertex; and the boosted decision tree (BDT) discrimination scores for both S1 and S2, which quantify the consistency of an event's waveform with a genuine recoil signal. Following the analysis performed by the XENONnT collaboration~\cite{XENON:2026ydt}, these observables are simultaneously accounted for in a full four-dimensional (4D) binned analysis with 81 bins per SR, yielding 243 bins in total. Unlike Refs.~\cite{DeRomeri:2024iaw,DeRomeri:2024hvc}, we no longer rely on correction factors nor on other factors such as the secondary scintillation gain $g_2$, the detection efficiency, or the charge yield. Instead, we calibrate directly against the multidimensional response templates provided in the XENONnT \cevns~data release~\cite{XENONnT_cevns_data_release,XENONnT_github}, which already incorporate these detector-response effects. We provide more details on our XENONnT simulation in Appendix~\ref{app:XENONnT_analysis}.

The analysis of the paired prompt scintillation (S1) and delayed electroluminescence (S2) signals in LZ data~\cite{LZ:2025igz} follows a similar approach to  PandaX-4T. The differential event rate is defined as
\begin{equation}
    \frac{\d R}{\d T_\mathcal{N}} =  \mathscr{E}(T_\mathcal{N})\int_{E_\nu^\text{min}}^{E_\nu^\text{max}} \d E_\nu \sum_{\ell=e,\mu,\tau}\frac{\d\phi_\ell}{\d E_\nu} \frac{\d\sigma_{\nu_\ell \mathcal{N}}}{\d T_\mathcal{N}}\,,
    \label{eq:diff_rate}
\end{equation}
where $\frac{\d\phi_\ell}{\d E_\nu}$ is the (oscillated) $\nu_\ell$ flux at the detector, $\frac{\d \sigma_{\nu_\ell \mathcal{N}}}{\d T_\mathcal{N}}$ the {\cevns} cross section, and $\mathscr{E}(T_\mathcal{N})$ the detector efficiency provided by the LZ Collaboration \cite{LZ:2025igz}. 
The lower integration limit depends on the nuclear recoil energy and reads as $E_\nu^\text{min}=\sqrt{m_\mathcal{N} T_\mathcal{N}/2}$, while $E_\nu^\text{max}$ is defined by the endpoint of the neutrino energy distribution, $E_\nu^\text{max}\approx 16$~MeV in the case of $^8$B neutrinos.
LZ data are presented in bins of S2 \cite{LZ:2025igz}; the number of events per bin $i$ is given by
\begin{equation}
    R_i = \mathcal{E} \int_i \frac{\d R}{\d n} \d n\,,
    \label{eq:diff_rate_n}
\end{equation}
where $\mathcal{E} = 5.7 $ t$\times$yr is the exposure achieved so far by the LZ experiment.
Within the experimental region of interest (ROI) and following \cite{LZ:2025igz}, we consider $5$ bins between $[155, 645]$ photons detected (phd), which correspond to nuclear recoil energies of $[1, 6]$~keV.
The differential event rate in Eq.~(\ref{eq:diff_rate_n}) is then expressed via a change of variables
\begin{equation}
    \frac{\d R}{\d n} = \frac{\d R}{\d T_{\mathcal{N}}} \frac{\d T_{\mathcal{N}}}{\d n}\,,
    \label{eq:change_variables}
\end{equation}
where the translation between the nuclear recoil energy and the S2 signal is given by
\begin{equation}
    n \equiv \mathrm{S2} = g_2T_{\mathcal{N}} Q_y(T_{\mathcal{N}})\,,
    \label{eq:nS2}
\end{equation}
with $g_2 = 34.0$ phd/electron, and the charge yield $Q_y(T_{\mathcal{N}})$ taken from \cite{LZ:2025igz}.

The total number of predicted events for LZ in each bin $i$ is given by
\begin{equation}
    N_i = R_i(1+\alpha) + B_i^{\rm AC}(1+\beta)\,,
    \label{eq:total_in_bin}
\end{equation}
where the spectrum of the dominant accidental coincidence (AC) background, $B_i^{\rm AC}$, is provided by the Collaboration \cite{LZ:2025igz}.
Through a combined analysis of both ionization and scintillation signals, LZ has observed 19 events in the ROI. We compare our predictions with the measured data $D_i$ using
\begin{equation}
    \chi^2 = \min\limits_{\alpha,\beta} \Bigg\{ 2 \Bigg[ \sum_i N_i - D_i + D_i  ~\text{ln}\Bigg( \frac{D_i}{N_i} \Bigg) \Bigg] + \Bigg( \frac{\alpha}{\sigma_\alpha} \Bigg)^2 + \Bigg( \frac{\beta}{\sigma_{\beta}} \Bigg)^2  \Bigg\}\,.
    \label{eq:chi2_analysis}
\end{equation}
Here, $\alpha$ ($\beta$) is a nuisance parameter with $\sigma_\alpha = 12\%$ ($\sigma_\beta = 30\%$) accounting for the $^8$B flux prediction (the normalization of the AC background component).

\section{Results}
\label{sec:results}
In the following we discuss the results of our analyses of SM physics in Sec.~\ref{sec:SM_physics}, neutrino NSI in Sec.~\ref{sec:NSI_physics}, and new light vector mediators in Sec.~\ref{sec:mediators_physics}. 

\subsection{SM physics: $^8$B flux and weak mixing angle}
\label{sec:SM_physics}

\begin{figure}
    \centering
    \includegraphics[width = 0.41 \textwidth]{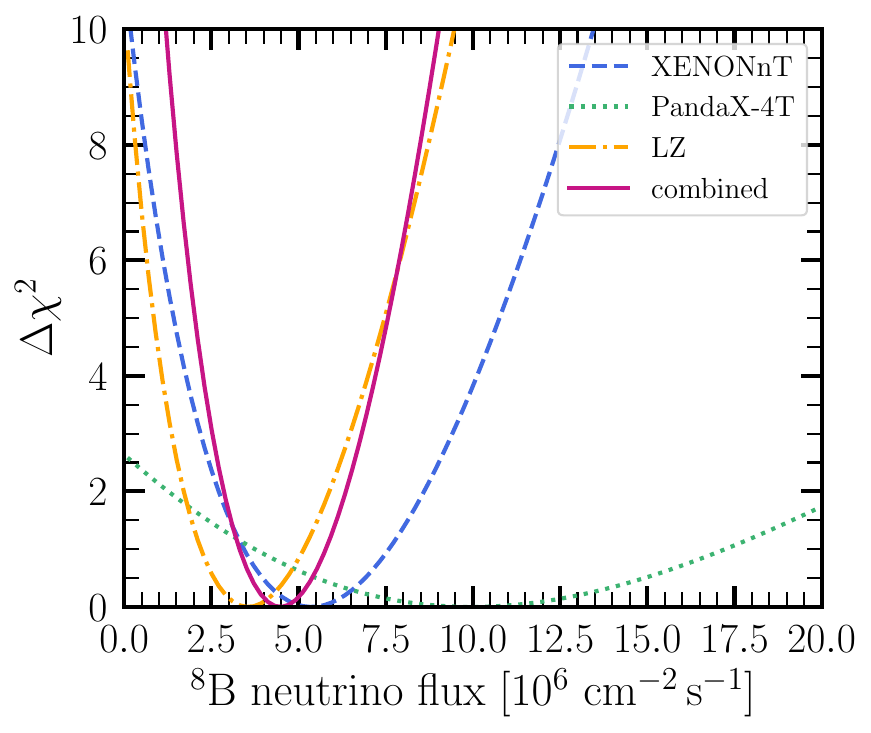}
    \includegraphics[width = 0.58\textwidth]{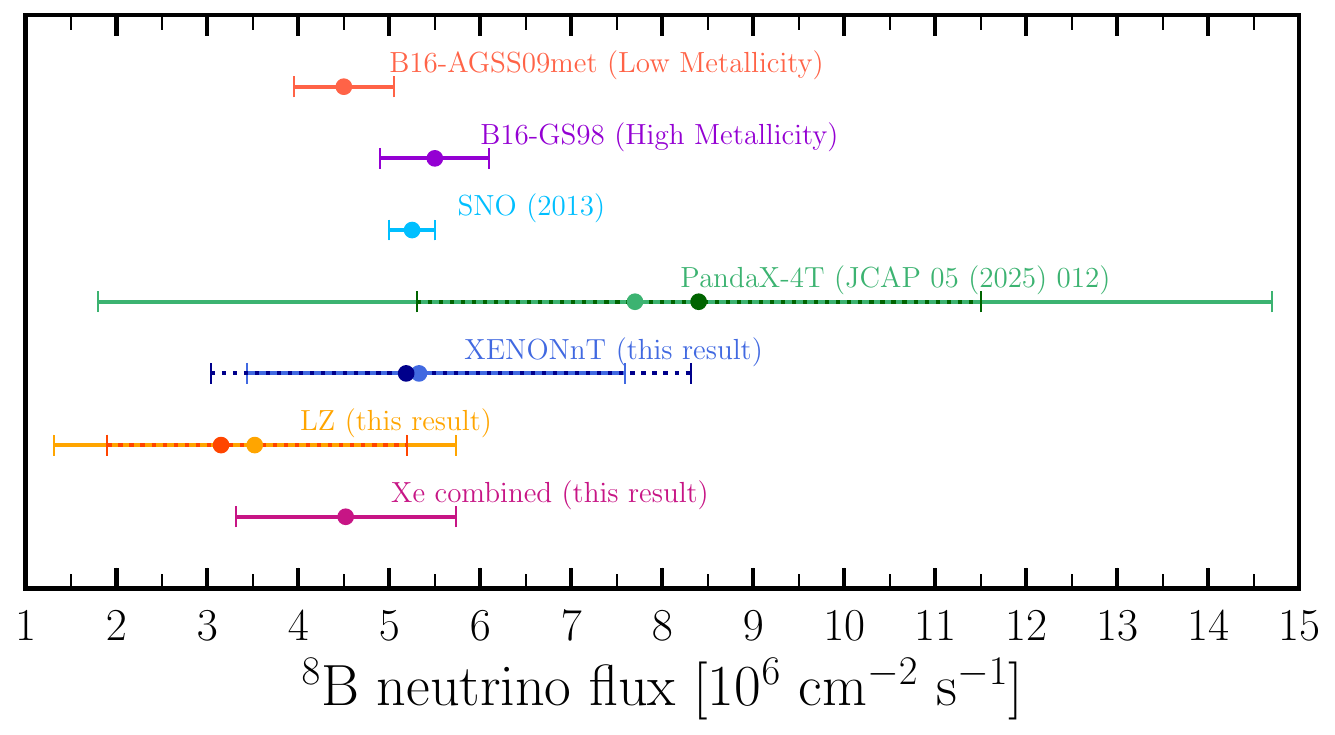}
        \caption{{\bf Left panel}: The $\Delta\chi^2$ profiles for the determination of the $^8$B solar neutrino flux for PandaX-4T (dotted green), XENONnT (dashed blue), LZ (dot-dashed orange), and the combined analysis (solid magenta). 
     {\bf Right panel}: $^8$B solar neutrino flux determination from our individual and combined analyses of solar \cevns~data, in comparison with other experimental measurements from neutral-current reactions \cite{Baxter:2021pqo,SNO:2011hxd,XENON:2024ijk,PandaX:2024muv,LZ:2025igz}, as well as predictions from low- and high-metallicity solar models~\cite{Vinyoles:2016djt}. 
     The superimposed dotted bars represent the official flux determinations as reported by each Collaboration.}
        \label{fig:8B_flux}
\end{figure}

The recent \cevns~measurements from DM direct detection experiments provide a simultaneous test of the prediction for the $^8$B neutrino flux from Standard Solar Models (SSMs) and of the SM prediction for the weak mixing angle, $\sin^2\theta_W$.

Regarding the determination of the $^8$B flux, the three Collaborations report results consistent with the predictions from SSMs and with previous dedicated solar neutrino experiments. At $68\%$ confidence level (CL), they obtain  $\Phi_\nu^{^8\text{B}} = (5.2^{+3.1}_{-2.1}~ , 8.4^{+3.1}_{-3.1} ~, 3.1^{+2.1}_{-1.3}) \times 10^6$ cm$^{-2}$ s$^{-1}$ from XENONnT, PandaX-4T and LZ data, respectively.
In deriving our results, we assume the SM prediction for the \cevns~cross section, fixing $\sin^2\theta_W = 0.23857$~\cite{ParticleDataGroup:2024cfk}.
The corresponding $\Delta\chi^2$ profiles are displayed in the left panel of Fig.~\ref{fig:8B_flux}. The green dotted and blue dashed curves show the PandaX-4T result previously presented in Ref.~\cite{DeRomeri:2024iaw} and the newly updated XENONnT result, respectively (see also the complementary analysis in~\cite{AtzoriCorona:2025gyz}), while the  orange dot-dashed curve shows the new LZ analysis.
The solid magenta line determines their combined result. At $1\sigma$, our individual and combined analyses give
\begin{align}
    \Phi_\nu^{^8\text{B}} &= (7.7^{+7.0}_{-5.9}) \times 10^6 ~\text{cm}^{-2} \text{s}^{-1} \quad (\text{PandaX-4T})\,,\\
    \Phi_\nu^{^8\text{B}} &= (5.3^{+2.3}_{-1.9}) \times 10^6 ~\text{cm}^{-2} \text{s}^{-1} \quad (\text{XENONnT})\,,\\
    \Phi_\nu^{^8\text{B}} &= (3.5^{+2.2}_{-2.2}) \times 10^6 ~\text{cm}^{-2} \text{s}^{-1} \quad (\text{LZ})\,,\\
    \Phi_\nu^{^8\text{B}} &= (4.5^{+1.2}_{-1.2}) \times 10^6 ~\text{cm}^{-2} \text{s}^{-1} \quad (\text{combined})\,.
\end{align}
We report these determinations in the right panel of Fig.~\ref{fig:8B_flux} as green, blue, orange, and magenta error bars, respectively. For completeness and ease of comparison, we also display as dotted error bars the official results quoted by each of the Collaborations~\rd{\cite{PandaX:2024muv,XENON:2026ydt,LZ:2025igz}.} 
Our extracted flux normalizations are consistent with the measurement from \textsc{SNO}~\cite{SNO:2011hxd}, also shown in the same panel, as well as with the predictions of both low- and high-metallicity solar models~\cite{Vinyoles:2016djt}.

\begin{figure}
    \centering
    \includegraphics[width = 0.41 \textwidth]{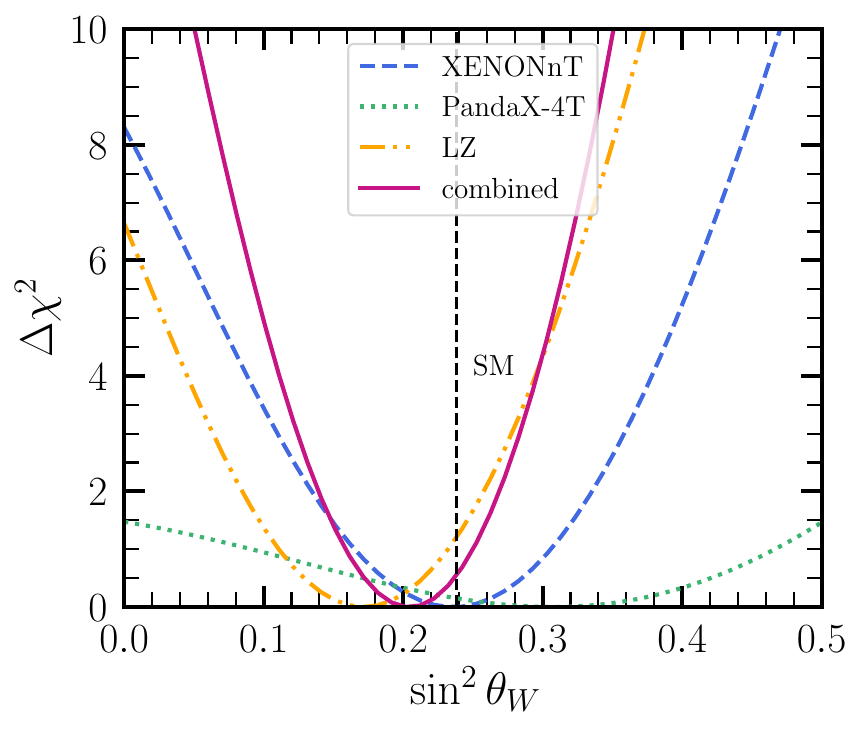}
    \includegraphics[width = 0.56\textwidth]{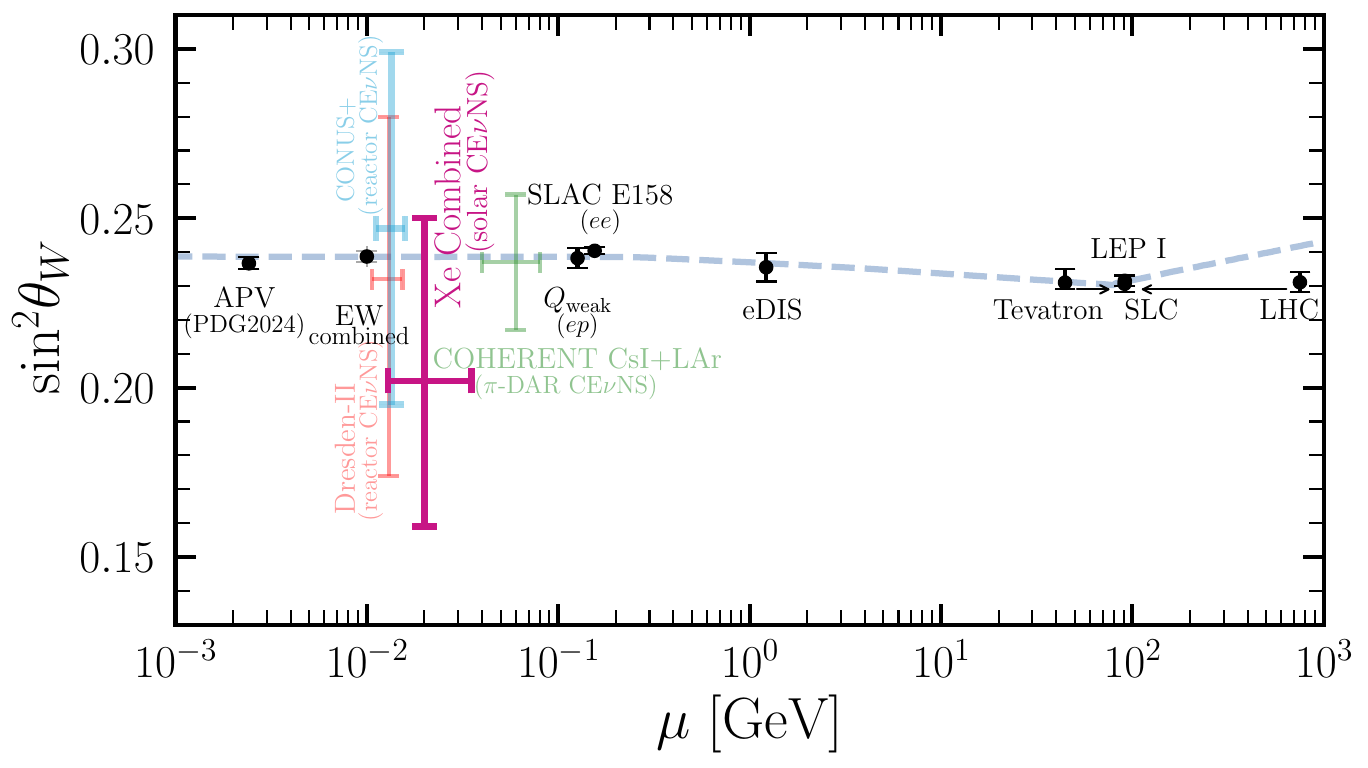}
        \caption{{\bf Left panel}: The $\Delta\chi^2$ profiles for the determination of the weak mixing angle for PandaX-4T (dotted green), XENONnT (dashed blue), LZ (dot-dashed orange), and the combined analysis (solid magenta). 
        The dashed vertical line corresponds to the  expected value of $\sin^2\theta_W$ within the SM. 
        {\bf Right panel}: Our $1\sigma$ combined {\cevns} measurement, together with the ones from other experiments at different energy scales~\cite{Majumdar:2022nby,DeRomeri:2022twg,NuTeV:2001whx,PVDIS:2014cmd,SLACE158:2005uay,Qweak:2018tjf,ParticleDataGroup:2024cfk,ALEPH:2005ab,CDF:2018cnj,ATLAS:2015ihy,ATLAS:2018gqq}. The running of the weak mixing angle as defined in the $\overline{\rm MS}$ renormalization scheme \cite{Erler:2017knj,Erler:2004in} is also shown (gray dashed line).}
        \label{fig:sw2_running}
\end{figure}

The observation of {\cevns} in DM direct detection experiments also enables a determination of the weak mixing angle, $\sin^2\theta_W$, at low-momentum transfer. In this case, we fix the neutrino flux normalization to $\Phi_\nu^{^8\text{B}} = 5.46 \times 10^6~\text{cm}^{-2} \text{s}^{-1}$~\cite{Vinyoles:2016djt,Baxter:2021pqo} with a corresponding $12\%$ uncertainty to account for the residual flux normalization error.
In addition to the PandaX-4T result already presented in \cite{DeRomeri:2024iaw} and the newly updated XENONnT result (see also~\cite{Maity:2024aji,AtzoriCorona:2025gyz}), we extract the $1\sigma$ best-fit value for LZ and compute the combined determination from the three experiments.
The corresponding $\Delta\chi^2$ profiles are shown in the left panel of Fig.~\ref{fig:sw2_running}, with the same color code applied before.
Our $1 \sigma$ results give
\begin{align}
    \sin^2\theta_W &= 0.30^{+0.16}_{-0.21} \quad (\text{PandaX-4T})\,,\\
    \sin^2\theta_W &= 0.23^{+0.07}_{-0.07}  \quad (\text{XENONnT})\,,\\
    \sin^2\theta_W &= 0.17^{+0.06}_{-0.06}  \quad (\text{LZ})\,,\\
    \sin^2\theta_W &= 0.20^{+0.05}_{-0.04}  \quad (\text{combined})\,.
\end{align}
We highlight the improvement in our determination of both the $^8$B flux and $\sin^2\theta_W$ with the updated XENONnT analysis based on the recently released dataset~\cite{XENON:2026ydt,XENONnT_cevns_data_release}.
The right panel of Fig.~\ref{fig:sw2_running} shows the combined best-fit value together with its associated $1\sigma$ uncertainty band. The gray dashed curve represents the renormalization group evolution of $\sin^2\theta_W$ in the $\overline{\rm MS}$ scheme~\cite{Erler:2017knj,Erler:2004in}, while other experimental constraints at different energy scales are also indicated~\cite{Majumdar:2022nby,DeRomeri:2022twg,NuTeV:2001whx,PVDIS:2014cmd,SLACE158:2005uay,Qweak:2018tjf,ParticleDataGroup:2024cfk,ALEPH:2005ab,CDF:2018cnj,ATLAS:2015ihy,ATLAS:2018gqq}.
As can be seen, the determination from DM direct detection experiments is becoming competitive with those obtained from reactor experiments such as CONUS+~\cite{DeRomeri:2025csu} and Dresden-II~\cite{Majumdar:2022nby} (see also Ref.~\cite{AtzoriCorona:2022qrf}). A more precise constraint is achieved by the combined analysis of COHERENT CsI and LAr data~\cite{DeRomeri:2022twg}, although at a somewhat higher energy scale due to the different neutrino source. Overall, the observation of neutrinos via the \cevns~channel in DM direct detection experiments provides, as a by-product, a valuable test of a fundamental SM parameter.

\subsection{Neutrino NSI}
\label{sec:NSI_physics}

\begin{figure}
    \centering
    \includegraphics[width=0.48\linewidth]{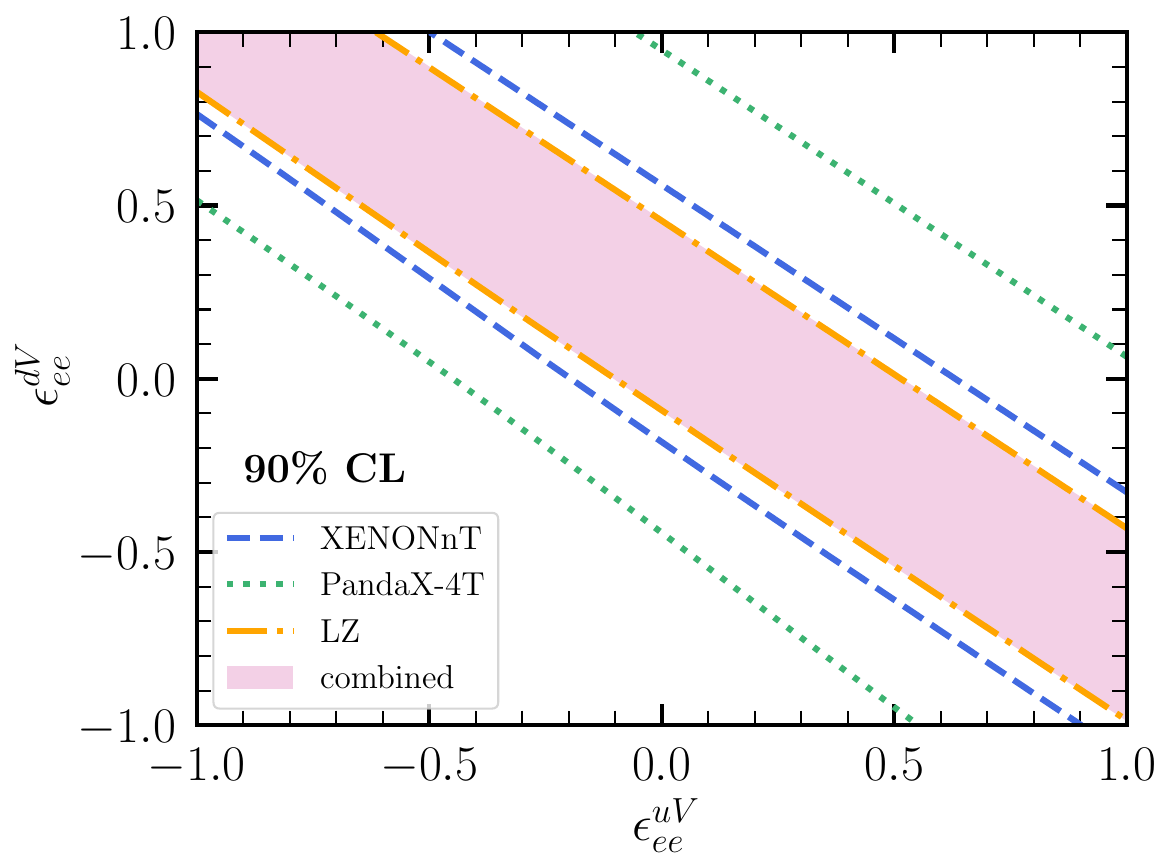}
        \includegraphics[width=0.48\linewidth]{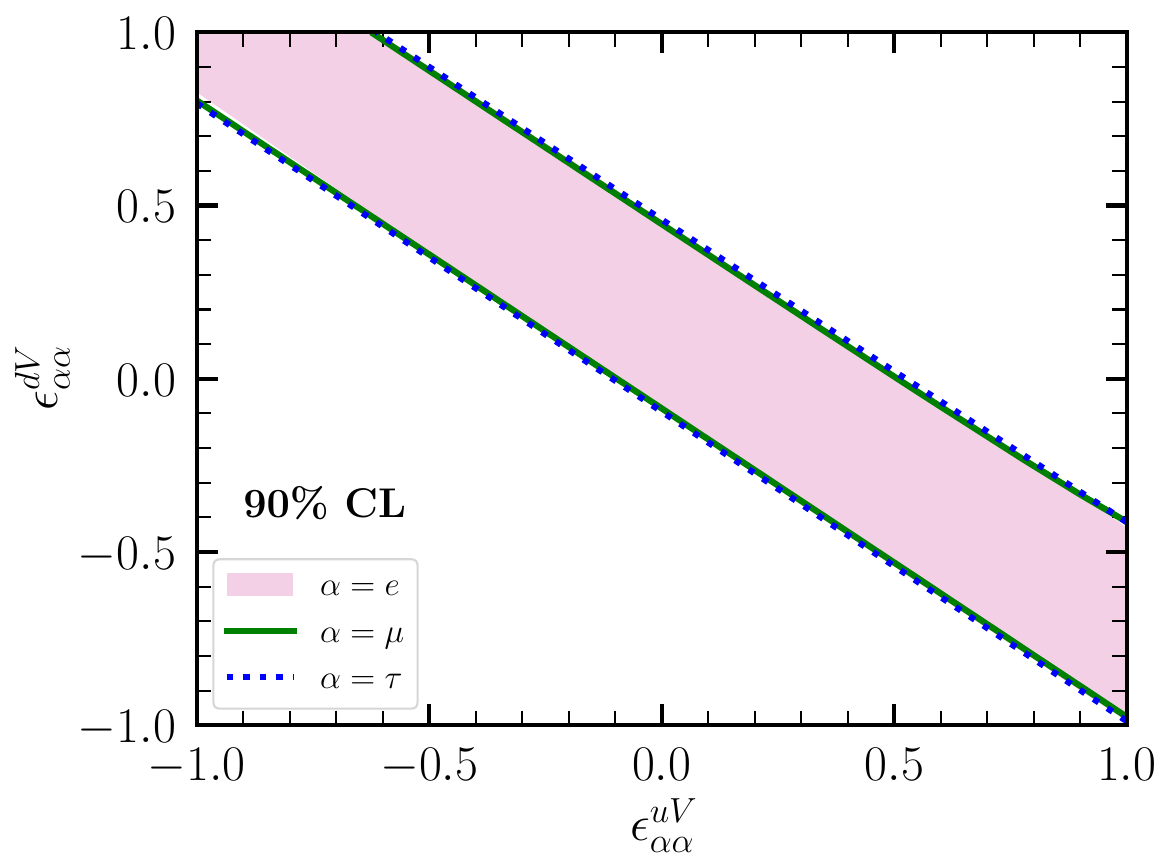}
    \caption{{\bf Left panel}: $90\%$ CL contours in the $(\epsilon_{ee}^{uV},\epsilon_{ee}^{dV})$ plane. All the other NSI parameters are fixed to zero. Green, blue and orange lines refer to PandaX-4T, XENONnT and LZ, respectively, while the pink region corresponds to their combined result. {\bf Right panel}: $90\%$ CL contours from the combined analysis in the diagonal $(\epsilon_{\alpha\alpha}^{uV},\epsilon_{\alpha\alpha}^{dV})$ plane, with $\alpha = e,\mu,\tau$ in shaded pink, green and blue color, respectively, and all the other NSI parameters fixed to zero.}
    \label{fig:NSI_ee}
\end{figure}

We present in this section the constraints on neutrino NSI obtained from the individual and combined analyses of XENONnT, PandaX-4T and LZ.
For simplicity, we first consider two nonzero NSI parameters at a time. 
We show in the left panel of Fig.~\ref{fig:NSI_ee} the $90\%$ CL contours in the $(\epsilon_{ee}^{uV},\epsilon_{ee}^{dV})$ parameter space. 
The results are illustrated for the individual and combined analyses, which turns out to be dominated by the LZ data. 
The constraints obtained from the combined XENONnT + PandaX-4T + LZ analysis are competitive with existing ones from COHERENT CsI+LAr~\cite{DeRomeri:2022twg} and CONUS+~\cite{DeRomeri:2025csu}.  
The right panel of Fig.~\ref{fig:NSI_ee} shows a comparison of the three NSI planes involving diagonal NSI couplings, $(\epsilon_{ee}^{uV},\epsilon_{ee}^{dV})$, $(\epsilon_{\mu \mu}^{uV},\epsilon_{\mu \mu}^{dV})$ and $(\epsilon_{\tau \tau}^{uV},\epsilon_{\tau \tau}^{dV})$. In this case, only the combined constraints are shown. 
The solar $^8$B neutrino-induced \cevns~data provide sensitivity to the $\epsilon_{\tau \tau}^{qV}$ NSI parameters, which are not accessible in measurements at spallation-source or reactor facilities.
The access to the $\epsilon_{\tau \tau}^{qV}$ channel makes DM direct detection facilities a great addition to dedicated {\cevns} experiments.

\begin{figure}
    \centering
    \includegraphics[width=0.48\linewidth]{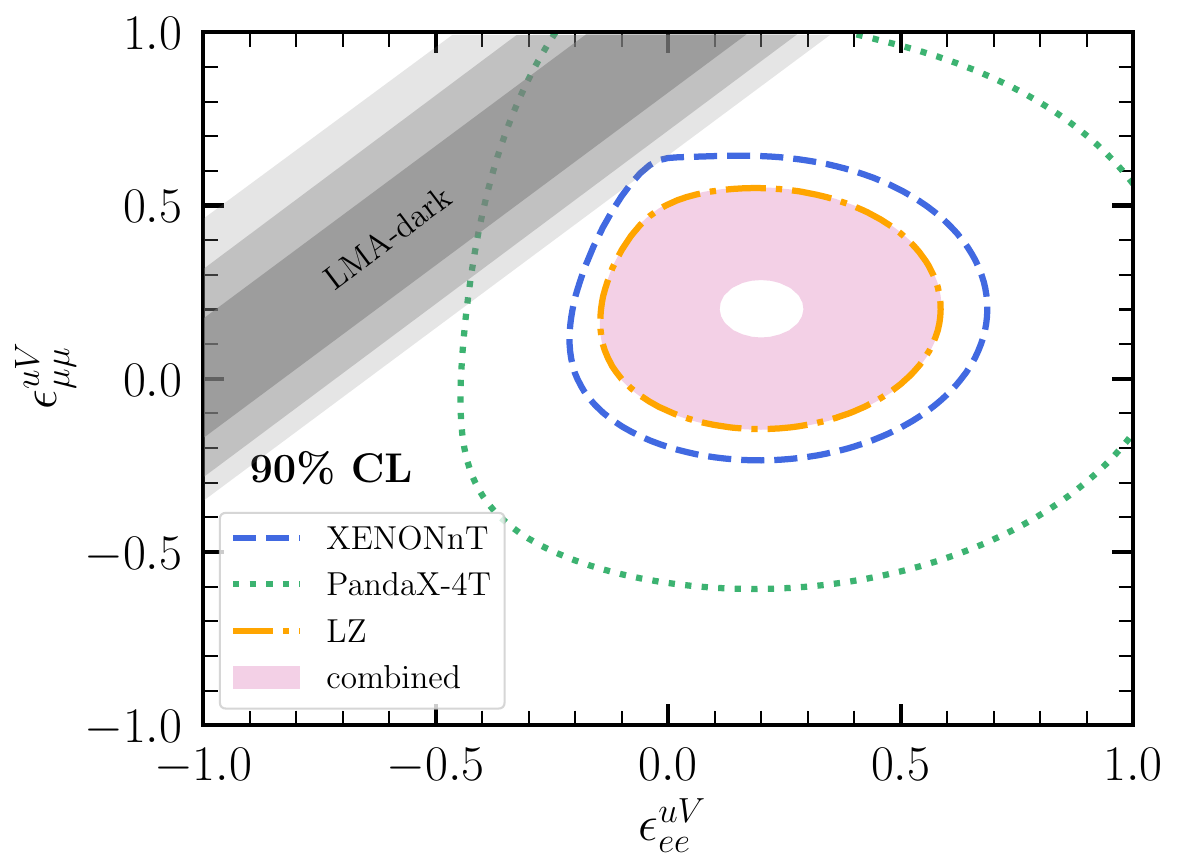}
        \includegraphics[width=0.48\linewidth]{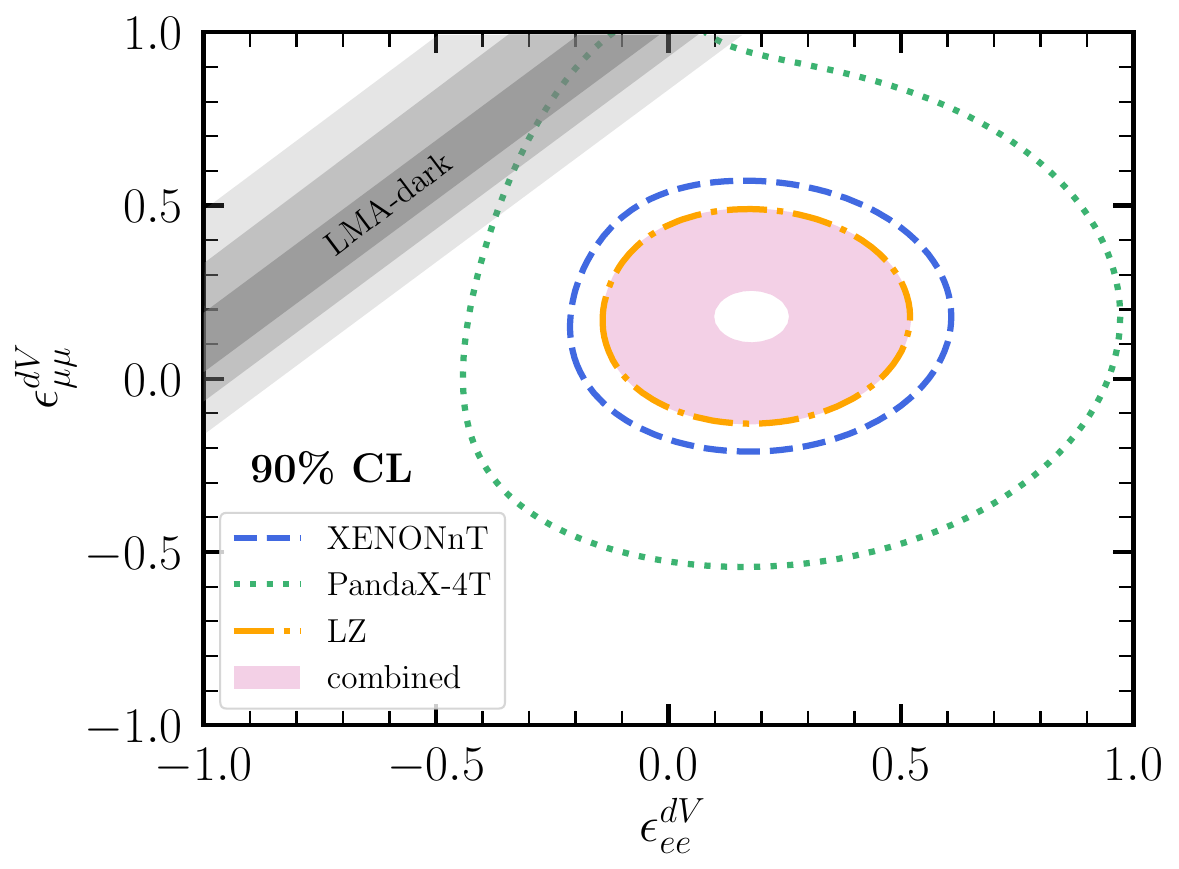}
    \caption{$90\%$ CL contours in the $(\epsilon_{ee}^{qV},\epsilon_{\mu\mu}^{qV})$ plane, with $q = u ~(d)$ in the {\bf left panel} ({\bf right panel}). Green, blue and orange lines refer to PandaX-4T, XENONnT and LZ, respectively, while the pink region corresponds to their combined result.
    We report here in gray also the parameter space region compatible with the LMA-dark solution.}
    \label{fig:NSI_ee_mm}
\end{figure}

From here on, we will consider NSI with either $u$- or $d$-quarks. 
In Fig.~\ref{fig:NSI_ee_mm} we show the $90\%$ CL bounds on the ($\varepsilon_{ee}^{uV}$, $\varepsilon_{\mu\mu}^{uV}$) (left panel) and the ($\varepsilon_{ee}^{dV}$, $\varepsilon_{\mu\mu}^{dV}$) (right panel) parameter space, while fixing all the other NSI parameters to zero.
Again, the combined analysis is driven by LZ, whose statistical power reduces the allowed parameter space, yielding a donut-shaped combined (pink) contour region. Note, however, that the SM solution, $\varepsilon_{ee}^{qV}$ = $\varepsilon_{\mu\mu}^{qV} = 0$ remains included within the shaded region, as expected.
In both panels, we also indicate the regions compatible with the so-called LMA-dark solution~\cite{Miranda:2004nb}, characterized by $\sin^2\theta_{12} > 0.5$ and sizeable NSI contributions. In the presence of sufficiently large NSI, the symmetry of the effective Hamiltonian under a simultaneous flip of the mass ordering and the $\theta_{12}$ octant can be restored through an appropriate sign change in the matter potential.
Assuming NSI with either $u$- or $d$-quarks only, our combined analysis excludes the parameter space required to realize the LMA-dark solution at more than $3\sigma$. This result is consistent with previous studies~\cite{Coloma:2017ncl,Chaves:2021pey,Gehrlein:2025isp}, which have highlighted the increasing tension of the LMA-dark scenario in light of recent \cevns\ data. We stress, however, that this conclusion does not necessarily extend to scenarios with light mediators, where the phenomenology can differ significantly~\cite{Denton:2018xmq,Denton:2022nol}.

\begin{figure}
    \centering
\includegraphics[width =0.8\textwidth]{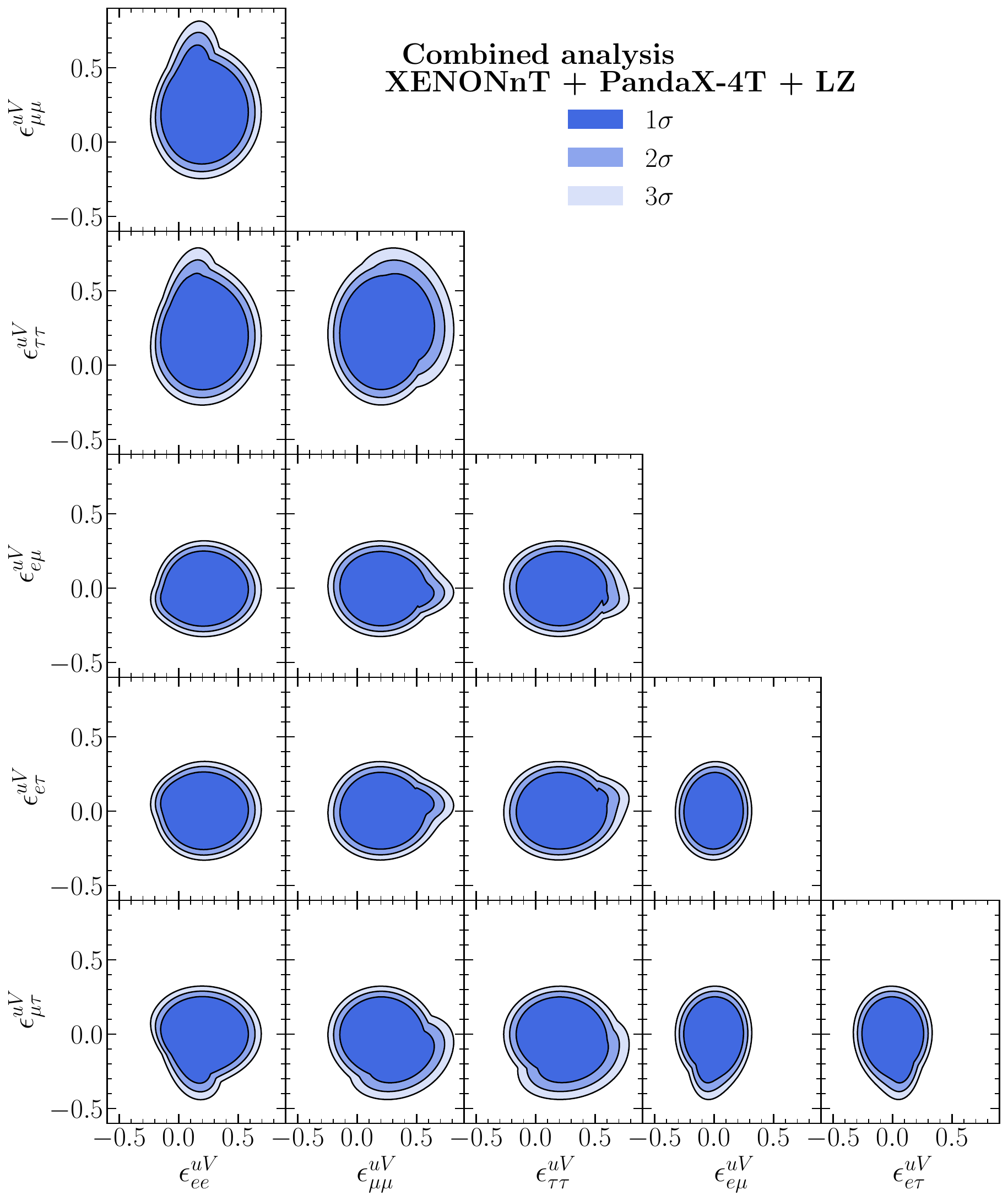}
\caption{The allowed regions at 1$\sigma$, 2$\sigma$ and 3$\sigma$ for 2 degrees of freedom for all pairs of NSI parameters with $u$-quarks from the combined XENONnT + PandaX-4T + LZ analysis. In each panel, we minimize with respect to all other NSI parameters that are not shown.}
    \label{fig:NSI_corner_u-quark}
\end{figure}

\begin{figure}
    \centering
\includegraphics[width =\textwidth]{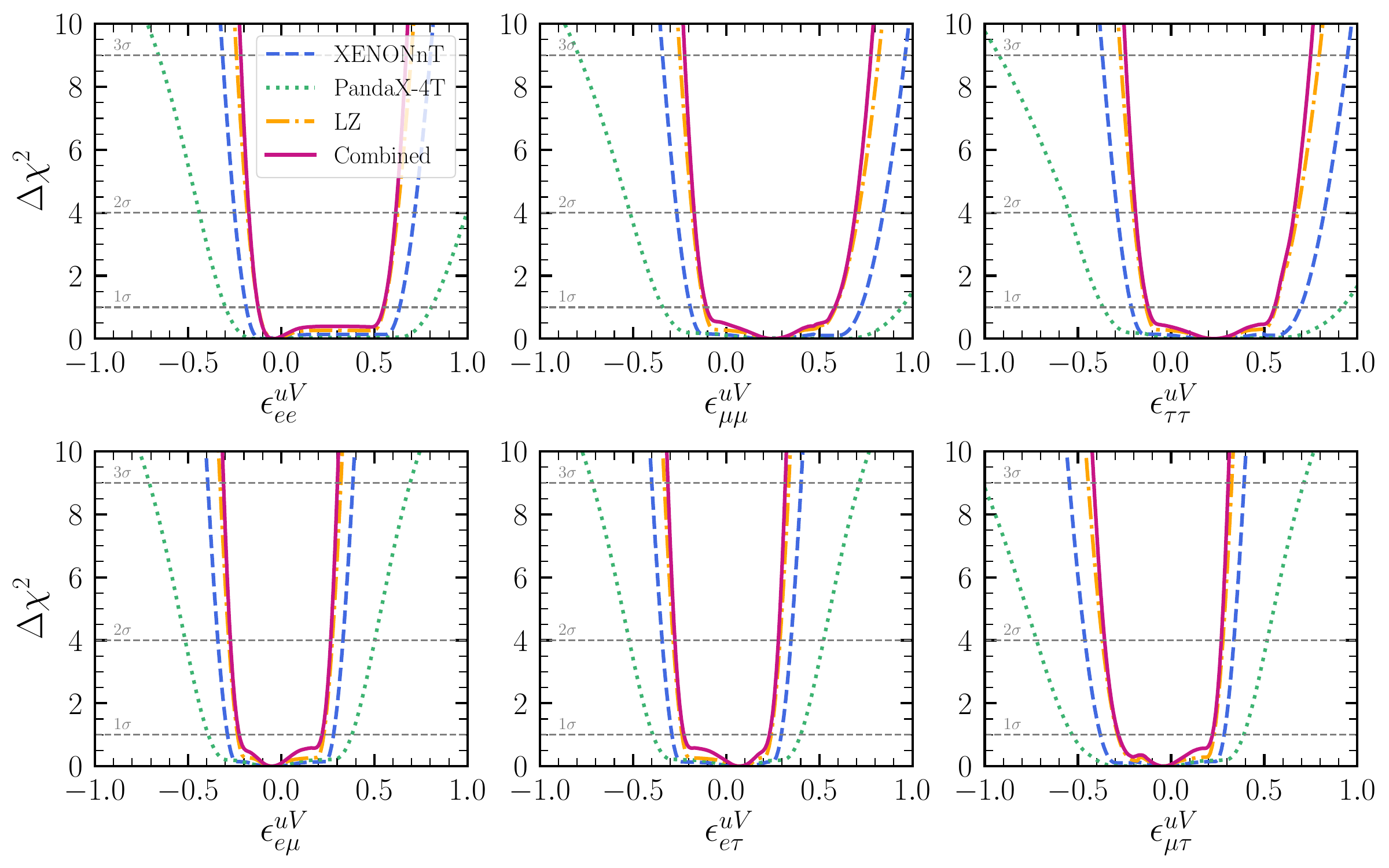}
\caption{$\Delta \chi^2$ profiles of all NSI parameters with $u$-quarks from the combined XENONnT + PandaX-4T + LZ analysis. The results for the individual analysis of each experiment are also depicted. In each panel, we minimize with respect to all other NSI parameters that are not shown. }
    \label{fig:NSI_1D_u-quark}
\end{figure}

Finally, we consider all the NSI as free parameters in the analysis and present marginalized constraints. 
Figure~\ref{fig:NSI_corner_u-quark} shows the result of our combined analysis of XENONnT, PandaX-4T and LZ data, assuming NSI with $u$-quarks only. 
Each panel displays a two-dimensional projection onto a pair of NSI parameters, with the allowed regions shown at $1\sigma$, $2\sigma$, and $3\sigma$ (from darker to lighter blue).
All remaining NSI parameters are profiled over in each projection.
The corresponding one-dimensional reduced $\chi^2$ profiles are shown in Fig.~\ref{fig:NSI_1D_u-quark} for each of the $\varepsilon_{\alpha\beta}^{uV}$. We present both the individual constraints from XENONnT, PandaX-4T, and LZ as well as the combined one. The corresponding allowed $1\sigma$ intervals are summarized in the upper half of Table~\ref{tab:NSI-quark}.
Note that we do not observe the double-minimum structure frequently reported in other similar works. This feature typically arises in analyses where only one NSI parameter is varied at a time. In contrast, when all NSI parameters and their correlations are consistently included, degeneracies arise, resulting in the relatively flat profiles displayed in the figure. We have verified  that, when allowing for only one NSI parameter at the time, the sensitivity improves and two minima become apparent, although their significance remain low. This behavior is also reflected in Fig.~\ref{fig:NSI_ee}, where a single allowed band is obtained instead of two separate bands. We understand this is a consequence of the low statistical significance of current solar \cevns~data and it can be improved with the accumulation of more statistics. 
The combined bounds derived here are mostly driven by the LZ result, and they are comparable in strength to those obtained from analyses of COHERENT data~\cite{Giunti:2019xpr,DeRomeri:2022twg,Coloma:2019mbs,Lozano:2025ekx}, although they remain weaker than the constraints achieved in global fits combining multiple dedicated neutrino experiments, in particular from oscillations~\cite{Coloma:2023ixt}.

\begin{table}[t!]
\centering
\begin{tabular}{|ccccc|}
\hline
Parameter & LZ & XENONnT & PandaX-4T & Combined \\
\hline
$\epsilon_{ee}^{uV}$      & [-0.123, 0.553] & [-0.188, 0.635] & [-0.300, 0.806] & [-0.123, 0.548] \\
$\epsilon_{\mu\mu}^{uV}$  & [-0.117, 0.588] & [-0.193, 0.724] & [-0.341, 0.943] & [-0.106, 0.583] \\
$\epsilon_{\tau\tau}^{uV}$ & [-0.134, 0.561] & [-0.213, 0.695] & [-0.366, 0.918] & [-0.123, 0.555] \\
$\epsilon_{e\mu}^{uV}$    & [-0.234, 0.229] & [-0.288, 0.281] & [-0.392, 0.380] & [-0.225, 0.219] \\
$\epsilon_{e\tau}^{uV}$   & [-0.237, 0.241] & [-0.291, 0.296] & [-0.393, 0.400] & [-0.226, 0.232] \\
$\epsilon_{\mu\tau}^{uV}$ & [-0.287, 0.231] & [-0.380, 0.285] & [-0.531, 0.390] & [-0.282, 0.222] \\
\hline
$\epsilon_{ee}^{dV}$      & [-0.116, 0.496] & [-0.175, 0.565] & [-0.276, 0.714] & [-0.116, 0.494] \\
$\epsilon_{\mu\mu}^{dV}$  & [-0.110, 0.472] & [-0.174, 0.540] & [-0.310, 0.750] & [-0.106, 0.470] \\
$\epsilon_{\tau\tau}^{dV}$ & [-0.126, 0.494] & [-0.193, 0.568] & [-0.335, 0.749] & [-0.122, 0.491] \\
$\epsilon_{e\mu}^{dV}$    & [-0.213, 0.208] & [-0.260, 0.253] & [-0.356, 0.344] & [-0.210, 0.205] \\
$\epsilon_{e\tau}^{dV}$   & [-0.215, 0.220] & [-0.262, 0.268] & [-0.356, 0.367] & [-0.212, 0.217] \\
$\epsilon_{\mu\tau}^{dV}$ & [-0.217, 0.210] & [-0.272, 0.255] & [-0.425, 0.353] & [-0.216, 0.207] \\
\hline
\end{tabular}
\caption{The $1\sigma$ limits for NSI with $u$-quarks and $d$-quarks obtained in this analysis. The combined limits are obtained by summing the individually profiled $\chi^2$ functions from the LZ, XENONnT, and PandaX-4T analyses.}
\label{tab:NSI-quark}
\end{table}

\begin{figure}
    \centering
\includegraphics[width =0.8\textwidth]{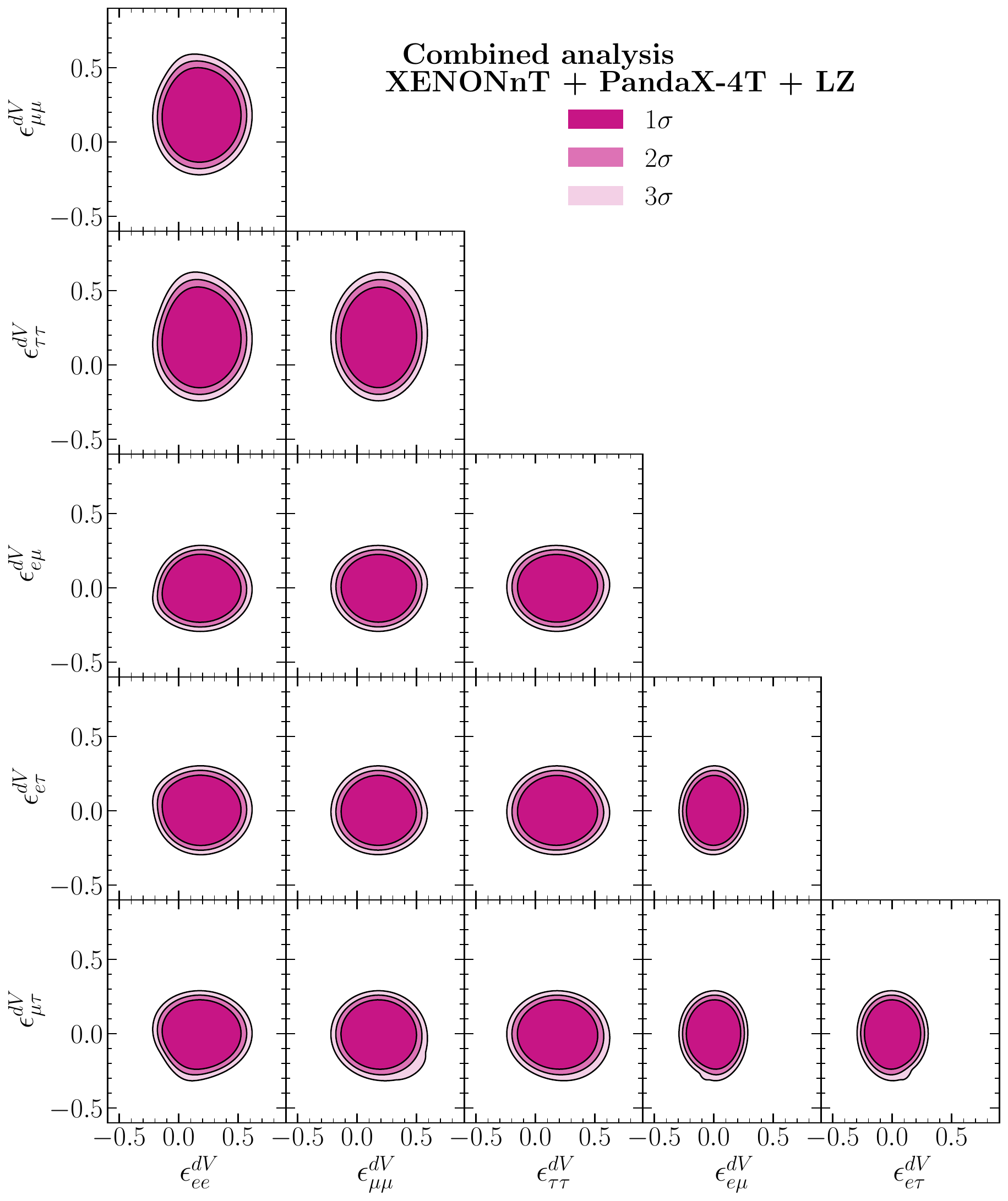}
\caption{Same as Fig.~\ref{fig:NSI_corner_u-quark}, but for NSI with $d$-quarks.}
    \label{fig:NSI_corner_d-quark}
\end{figure}

\begin{figure}
    \centering
\includegraphics[width =\textwidth]{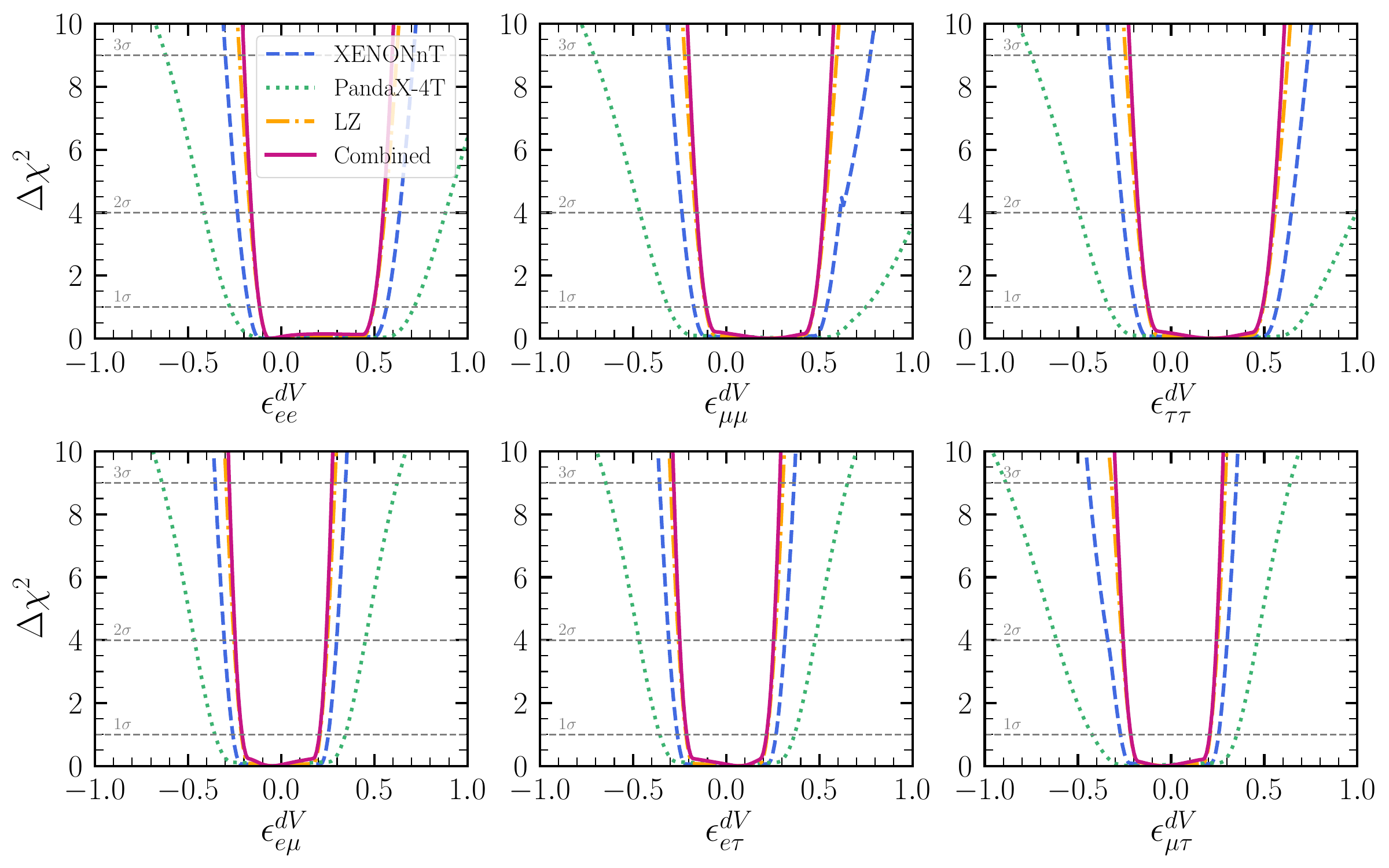}
\caption{Same as Fig.~\ref{fig:NSI_1D_u-quark}, but for NSI with $d$-quarks. }
    \label{fig:NSI_1D_d-quark}
\end{figure}

The results of the analogous analysis assuming NSI with $d$-quarks are presented in Figs.~\ref{fig:NSI_corner_d-quark} and~\ref{fig:NSI_1D_d-quark}, with the corresponding $1\sigma$ intervals summarized in the lower half of Table~\ref{tab:NSI-quark}. The qualitative conclusions closely mirror those obtained in the $u$-quark case: the allowed regions exhibit similar structures, and a flat profile appears once all parameter correlations are taken into account.
Since for xenon the number of neutrons exceeds the number of protons in the targets, the bounds on NSI with $d$-quarks are in general slightly more stringent than those with $u$-quarks. This is also manifested in less pronounced contour shapes, as visible in Fig.~\ref{fig:NSI_corner_d-quark}. Moreover, while our results confirm and complement previous analyses of solar \cevns~data~\cite{AristizabalSierra:2024nwf,AtzoriCorona:2025gyz,Li:2024iij,Gehrlein:2025isp}, a direct comparison (even focusing on the XENONnT and PandaX-4T datasets only) is not straightforward. The main differences stem from our use of a full spectral analysis and from the fact that we simultaneously vary and profile over all relevant NSI parameters, consistently accounting for their correlations. In addition, a key novelty of our work is the inclusion of the recent LZ data, which strengthens and updates the overall constraints.

\subsection{Light vector mediators}
\label{sec:mediators_physics}

\begin{figure}
    \centering
    \includegraphics[width = 0.49 \textwidth]{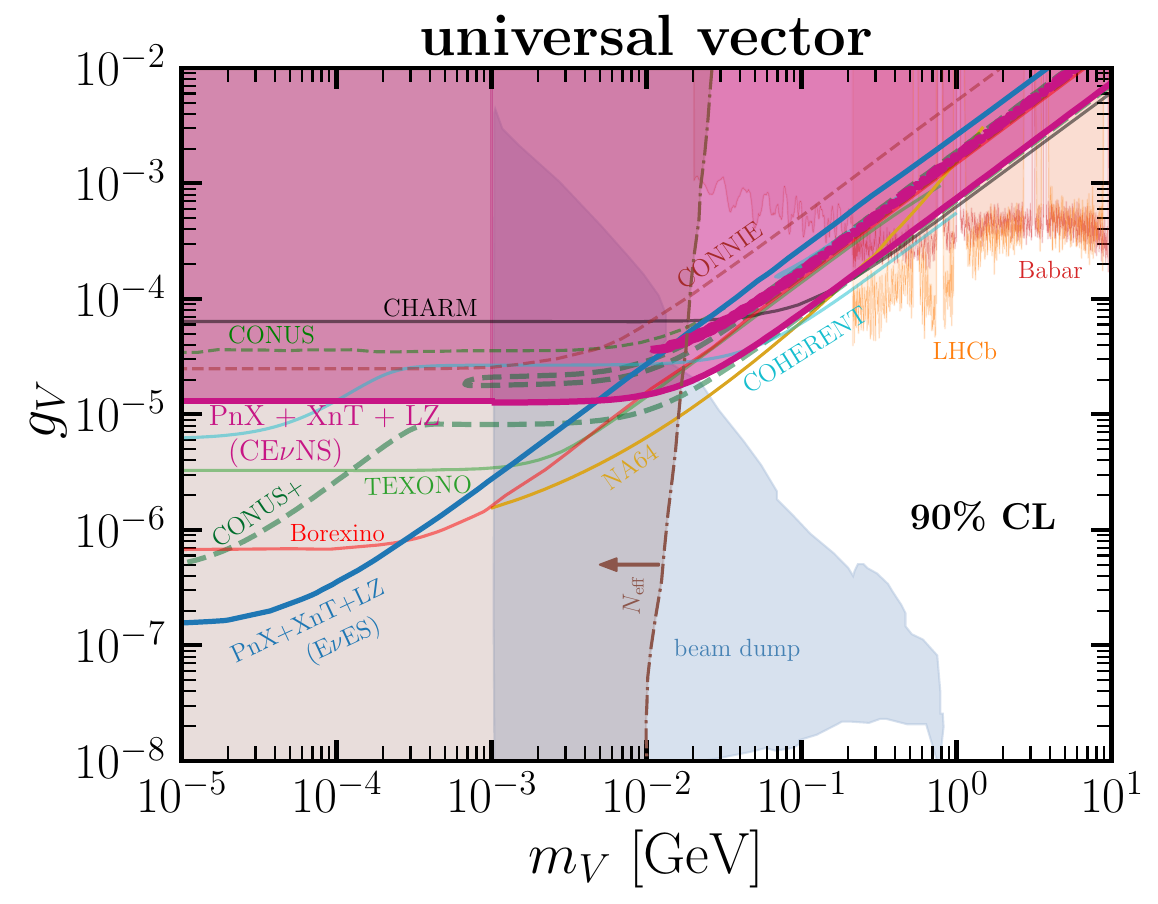}
\includegraphics[width = 0.49\textwidth]{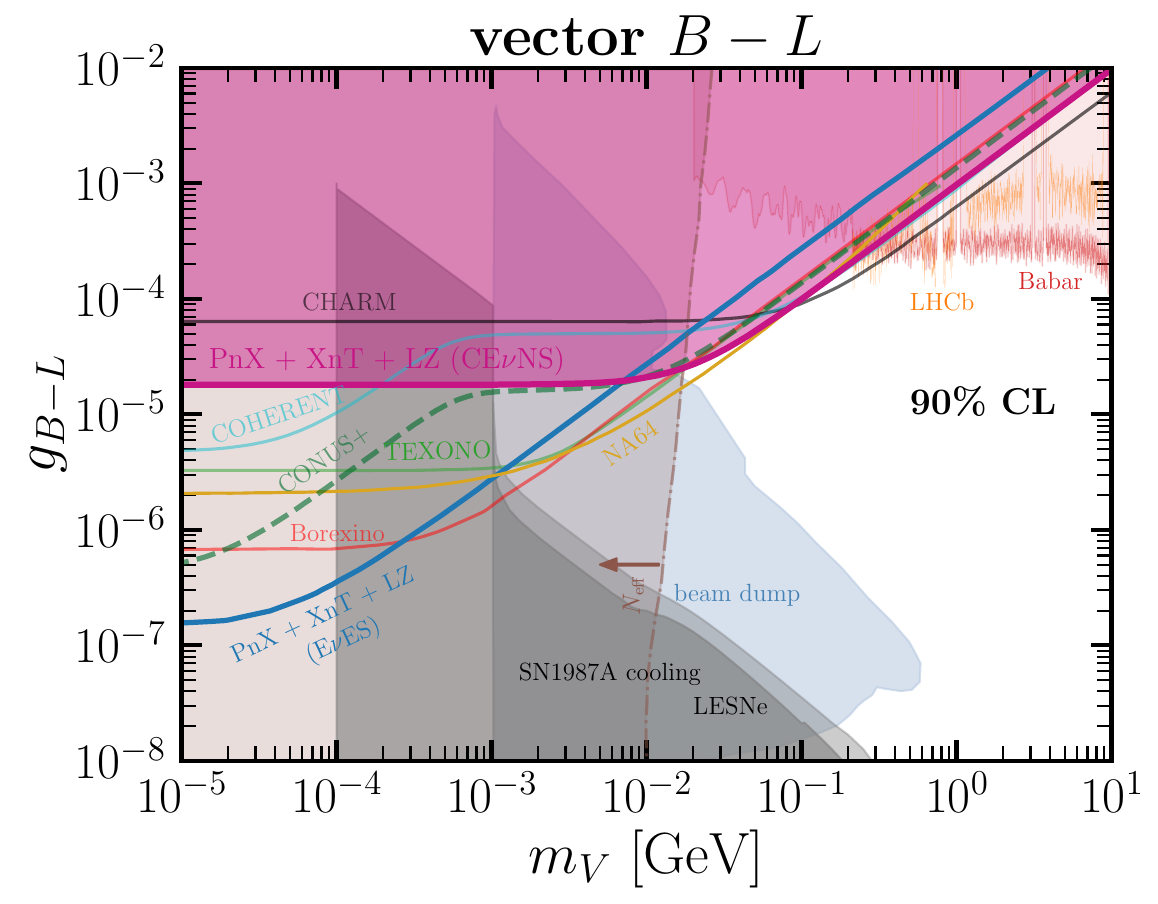}
        \caption{$90\%$ CL excluded regions (magenta shaded areas) from combined {\cevns} solar data (XENONnT + PandaX-4T + LZ) for universal vector ({\bf left panel}) and vector $B-L$ ({\bf right panel}) interactions. Existing bounds from other searches are shown for comparison.}
        \label{fig:light_mediators_combined}
\end{figure}

We finally discuss the constraints obtained from our combined analysis in the presence of light vector mediators.
We show our results in Fig.~\ref{fig:light_mediators_combined}, comparing them with other existing constraints. We report in magenta our $90\%$ CL constraints of the combined XENONnT + PandaX-4T + LZ {\cevns} analysis for the vector universal (left panel) and $B-L$ (right panel) mediators.
We also show existing constraints from other {\cevns} data from COHERENT~\cite{DeRomeri:2022twg,AtzoriCorona:2022moj}, CONNIE~\cite{CONNIE:2019xid,CONNIE:2024pwt},  CONUS~\cite{CONUS:2021dwh,Lindner:2024eng}, and CONUS+~\cite{DeRomeri:2025csu,Chattaraj:2025fvx,Ackermann:2025obx}.
Furthermore, we include limits from elastic neutrino-electron scattering data at BOREXINO~\cite{Coloma:2022avw}, CHARM-II~\cite{Bauer:2018onh}, and TEXONO~\cite{TEXONO:2009knm,Bauer:2018onh}, and from a combined analysis of PandaX-4T, XENONnT, and LZ electron recoil data~\cite{A:2022acy,DeRomeri:2024dbv}.
We further show exclusions from NA64~\cite{NA64:2021xzo,NA64:2022yly,NA64:2023wbi} and other fixed-target and beam-dump experiments, which include  E137~\cite{Bjorken:1988as}, E141~\cite{Riordan:1987aw}, KEK~\cite{Konaka:1986cb}, E774~\cite{Bross:1989mp}, Orsay~\cite{Davier:1989wz,Bjorken:2009mm,Andreas:2012mt}, and $\nu$-CAL~I~\cite{Blumlein:1990ay,Blumlein:1991xh,Blumlein:2011mv,Blumlein:2013cua}, CHARM~\cite{CHARM:1985anb,Gninenko:2012eq}, NOMAD~\cite{NOMAD:2001eyx}, PS191~\cite{Bernardi:1985ny,Gninenko:2011uv}, A1~\cite{Merkel:2014avp}, and APEX~\cite{APEX:2011dww}, and from colliders BaBar~\cite{BaBar:2014zli,BaBar:2017tiz} and LHCb~\cite{LHCb:2017trq}. Astrophysical and cosmological probes place important constraints on light mediators. While these limits depend strongly on the underlying model and would require a dedicated study for precise determination, we highlight with $N_\mathrm{eff}$ the parameter regions that could be incompatible with BBN~\cite{Esseili:2023ldf,Li:2023puz,Ghosh:2024cxi} and CMB~\cite{PhysRevD.110.075032} observations. Regions disfavored by supernova measurements are instead indicated by SN1987A and LESNe (for low-energy supernovae)~\cite{Caputo:2025avc,Heurtier:2016otg,Chang:2016ntp,Croon:2020lrf,Caputo:2021rux,Caputo:2022rca}.

We find that, in the universal vector scenario, the constraint is largely driven by XENONnT and therefore closely resembles the bound previously reported in Ref.~\cite{DeRomeri:2024iaw}.
In contrast, in the gauged $B-L$ framework, the sensitivity is dominated by the recent LZ data, even though we note an improvement of about a factor $\sim 1.4$ in the XENONnT bound at low mediator masses, due to the updated data analysis.
This behaviour can be traced back to the fact that the LZ \cevns~measurement lies slightly below the SM expectation. As a result, scenarios such as $B-L$, where the new contribution interferes constructively with the SM amplitude and enhances the event rate, are stronger constrained by LZ; See Appendix~\ref{Appendix} for more details.
In both models, the resulting bounds are comparable in strength to those obtained from other neutrino scattering experiments for mediator masses $m_V \gtrsim 10$ MeV. For lighter mediators, however, constraints from elastic neutrino–electron scattering become significantly more stringent. Our findings are consistent with previous studies, most notably Refs.~\cite{Blanco-Mas:2024ale,AtzoriCorona:2025gyz,Gehrlein:2025isp}, which also explored new light mediators in light of the XENONnT and PandaX-4T \cevns~data. In this respect, our analysis again extends the existing literature by incorporating the recent LZ nuclear-recoil results, thereby providing a comprehensive and up-to-date picture.

Within the universal vector scenario, it is also important to stress that an analysis based on a single dataset (XENONnT, LZ, or PandaX-4T alone) is not sufficient to lift the intrinsic parameter degeneracy. Individually, each experiment  allows for a narrow, unconstrained strip in the whole parameter space (with the exception of PandaX-4T which is presently not sensitive enough to resolve this strip, see, e.g., Ref.~\cite{DeRomeri:2024iaw} and Appendix~\ref{Appendix}) corresponding to an exact degeneracy in the total cross section, which cannot be excluded by that dataset alone. As detailed in Appendix~\ref{Appendix}, only the combined analysis of all three experiments is able to resolve this degeneracy and reduce the allowed region; eventually, only combining datasets with different target materials or improving recoil energy sensitivity in a single experiment would allow to fully probe it.
Remarkably, the resulting limits are already competitive with those obtained from dedicated \cevns~experiments such as COHERENT and CONUS+. The results shown here assume universal couplings of the light mediator to quarks and electrons, a well-motivated benchmark scenario. If this universality assumption is relaxed, the present bounds, together with other \cevns-based constraints, can become the dominant limits in the mediator mass range $m_V \simeq 20$–$200$ MeV.

\section{Conclusions}
\label{sec:concl}

The recent detection of $^8$B solar neutrinos via the \cevns~channel at DM direct detection facilities marks the entry of these experiments into the so-called \textit{neutrino fog}. On one hand, such solar neutrino signals slow down progress through the DM parameter space, the primary goal of these experiments. On the other hand, their detection via neutral-current nuclear scattering opens up new phenomenological research directions and enables the opportunity to employ current and future DM facilities as neutrino observatories.

Motivated by these recent experimental results, we build upon previous analyses to investigate their impact on both SM parameters and potential new physics. We perform a combined spectral analysis of \cevns~data collected by the XENONnT, PandaX-4T, and LZ experiments and extract a determination of the solar $^8$B neutrino flux normalization. Our results, both for individual datasets and the combined analysis, are in good agreement with the official values reported by the Collaborations, demonstrating the robustness of our statistical approach.

We further determine the weak mixing angle, a key SM parameter, at low energies, illustrating the power of DM facilities to perform precision SM tests and thereby significantly expanding their physics reach beyond their primary goal. Additionally, we explore the potential of these measurements to probe new neutrino interactions mediated by both heavy and light vector mediators. Our results show that DM direct detection experiments provide complementary constraints to dedicated neutrino facilities, yielding competitive bounds on nonstandard effective vector interactions as well as on the mass and coupling of light vector mediators, including universal and $B-L$ scenarios.

In conclusion, our work demonstrates that solar neutrino signals in DM detectors transform these experiments into competitive probes of both the SM and new physics, highlighting their evolving role as versatile observatories in astroparticle physics.

\section*{Acknowledgments}
We are grateful to the XENONnT \cevns~analysis team, in particular to Giovanni Volta and Dacheng Xu, for their guidance and support with the data release, as well as for many insightful conversations.
We also thank Mariam T\'ortola for helpful discussions. V.D.R. acknowledges financial support by the grant CIDEXG/2022/20 (from Generalitat Valenciana) and by the Spanish grants CNS2023-144124 (MCIN/AEI/10.13039/501100011033 and “Next Generation EU”/PRTR), PID2023-147306NB-I00, and CEX2023-001292-S (MCIU/AEI/10.13039/501100011033). D.K.P. acknowledges funding from the European Union’s Horizon Europe research and innovation programme under the Marie Skłodowska‑Curie Actions grant agreement No.~101198541 (neutrinoSPHERE).  G.S.G. has been supported by Sistema Nacional de Investigadoras e Investigadores (SNII, México) and by SECIHTI Project No. CBF-2025-I-1589.

\appendix

\section{Simulation of XENONnT signals}
\label{app:XENONnT_analysis}

In this Appendix we provide additional details regarding our simulation strategy for XENONnT. Following Ref.~\cite{XENON:2026ydt}, we compute the \cevns~signal assuming the following individual exposures: (1.174, 2.343, 3.250)~$\mathrm{t \times yr}$ in (SR0, SR1, SR2), respectively. For each SR, we compute 4D \cevns~rates assuming the four observables, namely the cS2, the quantile of $\mathrm{S2_{pre}/\Delta t_{pre}}$\footnote{$\mathrm{S2_{pre}}$ is the S2 area of the preceding high-energy event and $\mathrm{\Delta t_{pre}}$ is the time separation between that event and the isolated signal.}, the quantile of S1 BDT score and the quantile of S2 BDT score. For each observable we consider three bins as given by the XENONnT Collaboration, while the SR-dependent binning as well as the 4D response matrix mapping the nuclear recoil energy to the four observables of the experiment are taken from the XENONnT \cevns~data release~\cite{XENONnT_cevns_data_release}. The latter represents the probability of a given nuclear recoil energy to land in any of three SR-dependent bins for each observable. The 4D templates of the accidental coincidence (AC), neutron-recoil (NR) and electron-recoil (ER) backgrounds are also taken from~\cite{XENONnT_cevns_data_release}.

\begin{figure}
    \centering
    \includegraphics[width=\linewidth]{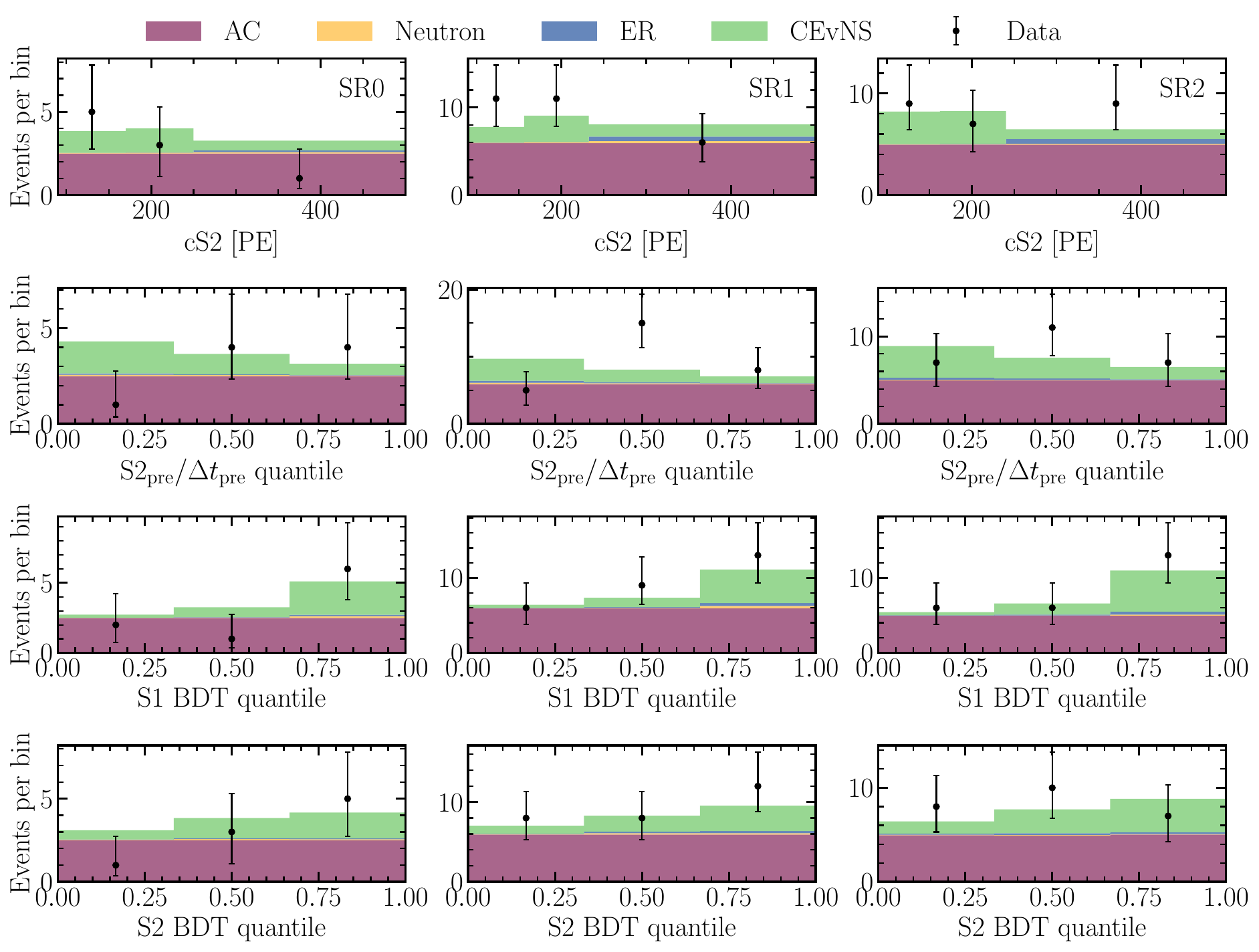}
    \caption{Our predicted \cevns~and background rates at XENONnT. The results are shown for the different SRs and projected on each of the four different observables: cS2, $\mathrm{S2_{pre}/\Delta t_{pre}}$, S1 BDT quantile and S2 BDT quantile.}
    \label{fig:XENONnT_simulation}
\end{figure}

Figure~\ref{fig:XENONnT_simulation} shows the our predicted \cevns~signal and background spectra, projected into each of the four observables for the different SRs, and is found to be in excellent agreement with Ref.~\cite{XENON:2026ydt}. As discussed in the main text, in our present analysis we consider the $^8$B flux normalization $\Phi_\nu^{^8\text{B}} = 5.46 \times 10^6~\text{cm}^{-2} \text{s}^{-1}$~\cite{Vinyoles:2016djt,Baxter:2021pqo} which leads to a total 16.9~\cevns~events. Instead, by considering $\Phi_\nu^{^8\text{B}} = 5.25 \times 10^6~\text{cm}^{-2} \text{s}^{-1}$~\cite{SNO:2011hxd} as done by XENONnT we estimate 16.2~\cevns~events, also in excellent agreement with the 16 expected events reported in Ref.~\cite{XENON:2026ydt}.

In this work, statistical inference relies on a 4D  Poisson $\chi^2$ analysis, consisting of 81 bins per SR (243 bins for the three SRs). Individual systematic uncertainties are introduced for the three AC backgrounds per SR, while the systematic uncertainties associated to ER and neutron backgrounds are shared among the SRs, all taken from~\cite{XENON:2026ydt}. Concerning the \cevns~signal, the corresponding systematic uncertainty accounts for a 12\% flux normalization uncertainty and a 5\% detector volume uncertainty.

\section{Individual bounds on light vector mediators}
\label{Appendix}

\begin{figure}[t!]
    \centering
\includegraphics[width = 0.48 \textwidth]{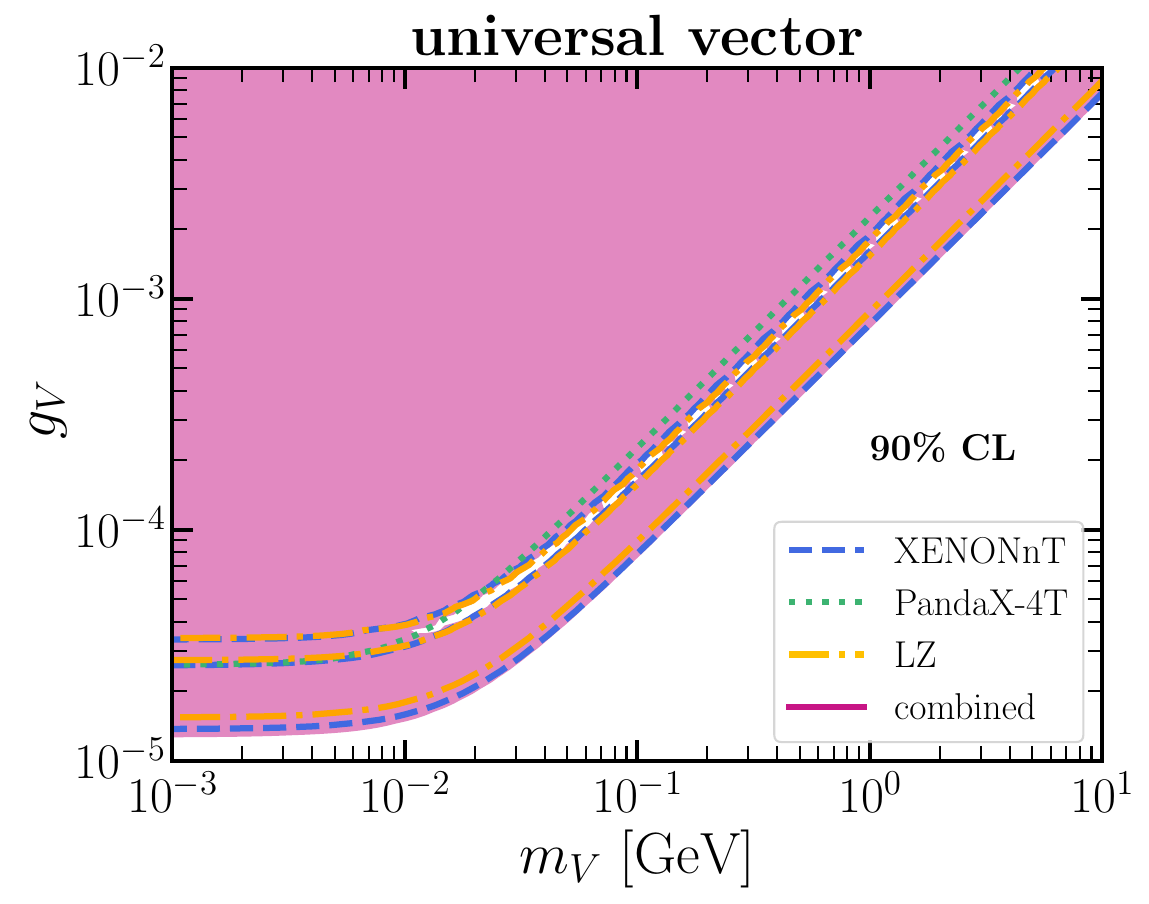}
\includegraphics[width = 0.48 \textwidth]{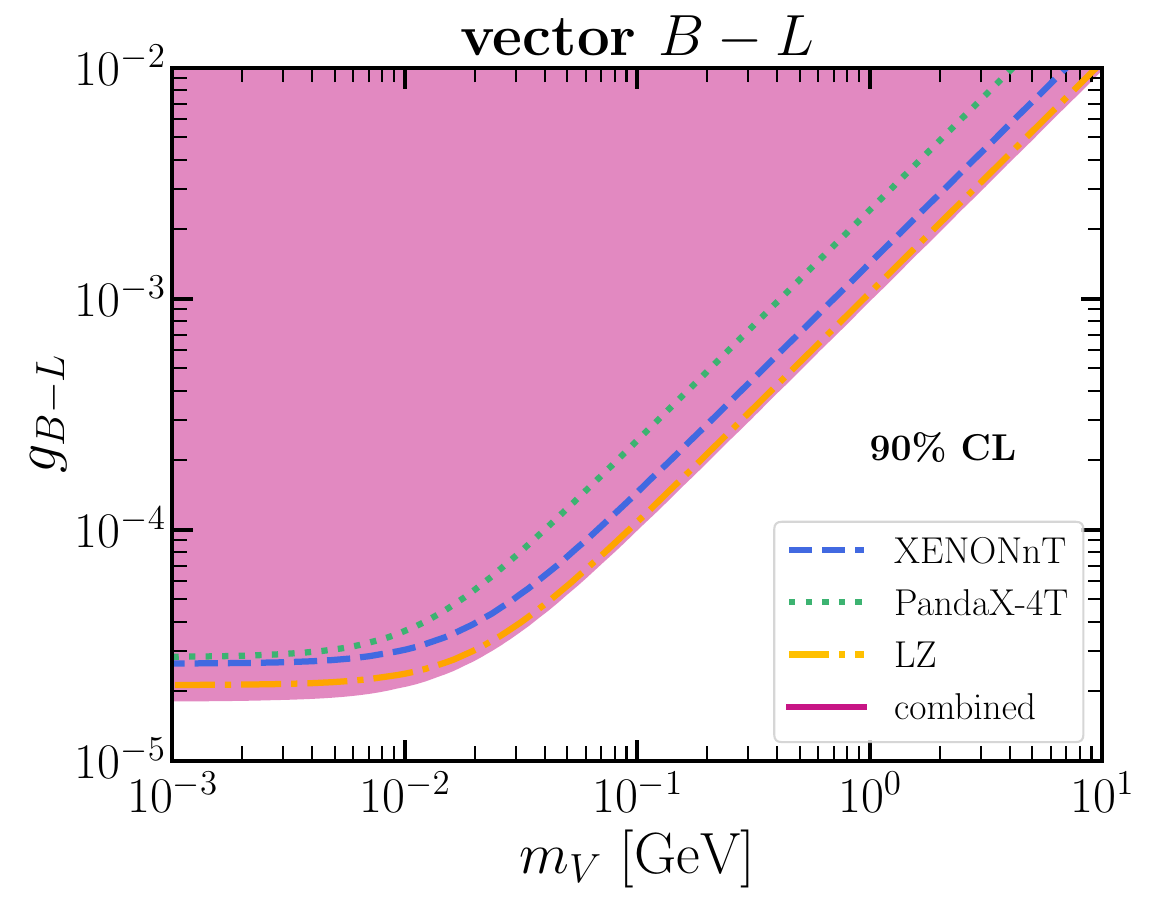}
        \caption{$90\%$ CL exclusion limits for universal ({\bf left panel}) and $B-L$ vector ({\bf right panel}) interactions. 
        Blue, green and orange lines correspond to XENONnT, PandaX-4T and LZ bounds, respectively, while the magenta refers to the combined exclusion region.}
        \label{fig:light_mediators_individual}
\end{figure}

The individual $90\%$ CL bounds in the mediator mass-coupling plane, obtained from each experiment are displayed in Fig.~\ref{fig:light_mediators_individual} for the universal (left panel) and $B-L$ (right panel) vector interactions. 
The improvement resulting from the new LZ data is particularly notable in the case of the $B-L$ model. The shapes of the constraints can be understood as follows: for light mediators ($m_V \ll \qtransfer$), a saturation of the depicted constrains happens for $g_V^2 = -\frac{2 \sqrt{2} G_F m_\mathcal{N} Q_{V, \ell}^\mathrm{SM} T_\mathcal{N}}{3 A \kappa}$, i.e. the constraint depends only on the coupling. 
On the other hand, in the heavy mediator regime ($m_V \gg \qtransfer$), the constraints appear as straight lines with positive slope and depend on the ratio $\frac{g_V^2}{m_V^2} = -\frac{\sqrt{2} G_F Q_{V, \ell}^\mathrm{SM}}{3 A \kappa}$. 
Moreover, in the case of the universal vector model, a full degeneracy of the cross section occurs for
\begin{equation}
    \frac{C_V}{\sqrt{2}G_F Q_{V,\ell}^\mathrm{SM}\left(m_{V}^2+2 m_\mathcal{N} T_\mathcal{N}\right)}=-2 \, .
    \label{eq:SM_cancellation}
\end{equation}
It is interesting to notice that an individual analysis of LZ, PandaX-4T or XENONnT data is not sufficient to break the degeneracy. 
However, this becomes possible via a combined analysis of all the available data performed in the present work, see e.g. the magenta exclusion regions, without necessarily the need of recurring to different target materials. The parameter space region where the destructive interference occurs is indeed dependent on the $N/Z$ ratio typical of the target material~\cite{AtzoriCorona:2022moj,Liao:2017uzy}.

\bibliographystyle{utphys}
\bibliography{bibliography}  

\end{document}